\documentclass[10pt,journal,compsoc]{IEEEtran}

\usepackage{xspace}
\xspaceaddexceptions{\%}
\xspaceremoveexception{-}
\usepackage{amsmath}
\usepackage{lstautogobble}
\usepackage{zi4}
\usepackage{amsfonts}
\usepackage{xr}
\usepackage{tabularx}
\usepackage{booktabs}
\usepackage{multirow}
\usepackage{listings}
\usepackage{multirow}
\usepackage{url}
\usepackage{hyperref}
\usepackage{booktabs}
\usepackage{color}
\usepackage{colortbl}
\usepackage{pgfplots}
\usepackage{tikz}
\usepackage{subfig}
\usepackage{adjustbox}
\usepackage{tikz}
\usepackage{qtree}
\usepackage{pgf-pie}
\usepackage{paralist}
\usepackage{filecontents}
\usepackage{graphicx}
\usepackage{pgfplots}
\usetikzlibrary{pgfplots.dateplot,positioning}
\usetikzlibrary{fadings}
\usepackage{import}
\usepackage{relsize}
\usepackage{float}
\usepackage{balance}
\usepackage[T1]{fontenc}
\usepackage{framed}
\usepackage{url}
\usepackage{listings}
\usepackage{balance}
\usepackage{boxedminipage}

\definecolor{color0}{rgb}{0.886274509803922,0.290196078431373,0.2}
\definecolor{color1}{rgb}{0.203921568627451,0.541176470588235,0.741176470588235}
\definecolor{color2}{rgb}{0.556862745098039,0.729411764705882,0.258823529411765}
\definecolor{color3}{rgb}{0.00392156862745098,0.466666666666667,0.756862745098039}
\definecolor{color4}{rgb}{0.456862745098039,0.429411764705882,0.658823529411765}

\definecolor{bluekeywords}{rgb}{0.13, 0.13, 1}
\definecolor{greencomments}{rgb}{0, 0.5, 0}
\definecolor{redstrings}{rgb}{0.9, 0, 0}
\definecolor{graynumbers}{rgb}{0.5, 0.5, 0.5}

\lstdefinestyle{diffs}{
  moredelim=**[is][\it\color{red}]{@}{@},
  moredelim=**[is][\it\color{green}]{H}{H},
}

\lstdefinestyle{customc}{
  belowcaptionskip=1\baselineskip,
  breaklines=true,
  frame=L,
  xleftmargin=\parindent,
  language=C,
  showstringspaces=false,
  basicstyle=\footnotesize\ttfamily,
  keywordstyle=\bfseries\color{green!40!black},
  commentstyle=\itshape\color{purple!40!black},
  identifierstyle=\color{blue},
  stringstyle=\color{orange},
}

\lstdefinestyle{algo}{
  mathescape=true,
  language=Python,
  literate=
  {<=}{$\leftarrow{}$}{1}
  {U=}{$\cup{}$= }{1}
  {-0}{$\emptyset{}$}{1}
  {==}{$={}$}{1},
  morekeywords={function,endfor,shuffle,remove_arid_nodes,changed_lines,generate_mutant,can_generate,order_by_historic_usefulness},
}

\newcommand{\hl}[2]{\relax}

\newif\ifTR
\TRtrue

\def\defTerm{\textit}

\def\prod{productive\xspace}

\def\cl{changelist\xspace}

\def\cls{changelists\xspace}

\def\PleaseFix{\textit{Please fix}\xspace}

\def\MTS{Mutation Testing Service\xspace}

\definecolor{lightgray}{gray}{0.9}
\def\excl{\relax}

\def\ballparkCLsTotal{760,000\xspace}
\def\ballparkMutantsGenerated{17 million\xspace}
\def\ballparkMutantsSurfaced{2 million\xspace}

\def\unprodStartPhaseOne{85\%\xspace}

\def\prodStartPhaseOne{15\%\xspace}
\def\prodEndPhaseOne{80\%\xspace}

\def\prodStartPhaseTwo{80\%\xspace}
\def\prodEndPhaseTwo{89\%\xspace}

\def\numMutantsGenerated{16,935,148\xspace}
\def\numMutantsSurfaced{2,110,489\xspace}

\def\percentMutantsKilled{87.5\%\xspace}
\def\numCLs{776,740\xspace}
\def\listPLs{C++, Java, Go, Python, TypeScript, JavaScript, Dart, SQL, Common Lisp, and Kotlin\xspace}
\def\listPLsExcl{SQL, Common Lisp, and Kotlin\xspace}
\def\numPLs{10\xspace}

\def\numMutantsFeedback{66,798\xspace} 

\def\numMutantsContext{4,068,241\xspace}

\def\percentMutantsPleaseFix{82\%\xspace} 

\def\ogap{-5pt}
\def\igap{-5pt}

\newcommand{\p}{$\scriptsize{\pm}$}

\def\<#1>{\codeid{#1}}
\newcommand{\codeid}[1]{\ifmmode{\mbox{\small\ttfamily{#1}}}\else{\small\ttfamily #1}\fi}
\newcommand{\codeidsmall}[1]{\ifmmode{\mbox{\smaller\ttfamily{#1}}}\else{\smaller\ttfamily #1}\fi}
\def\|#1|{\code{#1}}
\newcommand{\code}[1]{\ifmmode{\mbox{\ttfamily{#1}}}\else{\ttfamily #1}\fi}

\newcommand{\etal}{et al.\xspace}

\newcolumntype{H}{>{\setbox0=\hbox\bgroup}c<{\egroup}@{}}

\newenvironment{result}%
{\medskip
	\noindent
	\let\emph=\textbf
	\begin{boxedminipage}{\columnwidth}
      \em}%
		{
        \end{boxedminipage}%
	\medskip
}


\ifCLASSOPTIONcompsoc
  \usepackage[nocompress]{cite}
\else
  \usepackage{cite}
\fi
\ifCLASSINFOpdf
\else
\fi
\usepackage{fixltx2e}
\begin{document}
%
\title{Practical Mutation Testing at Scale \\ {\Large A view from Google}}
%
%
%
%

\author{Goran~Petrovi\'{c},
        Marko~Ivankovi\'{c},
        Gordon Fraser,
		Ren{\'e} Just

\IEEEcompsocitemizethanks{
\IEEEcompsocthanksitem Goran Petrovi\'{c} and Marko Ivankovi\'{c} are with Google LLC. \protect\\
E-mail: goranpetrovic@google.com, markoi@google.com
\IEEEcompsocthanksitem Gordon Fraser is with the University of Passau \protect\\
E-mail: gordon.fraser@uni-passau.de
\IEEEcompsocthanksitem Ren{\'e} Just is with the University of Washington \protect\\
E-mail: rjust@cs.washington.edu}%
\thanks{This work has been submitted to the IEEE for possible publication. Copyright may be transferred without notice, after which this version may no longer be
accessible.}
}

\IEEEtitleabstractindextext{%
\begin{abstract}
Mutation analysis assesses a test suite's adequacy by measuring its ability to detect small artificial
faults, systematically seeded into the tested program.
Mutation analysis is considered one of the strongest test-adequacy criteria.
Mutation testing builds on top of mutation analysis and is a testing technique that uses mutants as test goals to create or improve a test suite.
Mutation testing has long been considered intractable because the sheer number of mutants that can be created represents an
insurmountable problem---both in terms of human and computational effort. This has hindered the adoption of mutation testing as an industry standard.
For example, Google has a codebase of two billion lines of code and more than 500,000,000 tests are executed on a daily basis.
The traditional approach to mutation testing does not scale to such an environment; even existing solutions to speed up mutation analysis are insufficient to make it computationally feasible at such a scale.

To address these challenges, this paper presents a scalable approach to mutation testing
based on the following main ideas: (1) Mutation testing is done incrementally,
mutating only \emph{changed code} during code review, rather than the entire code
base; (2) Mutants are filtered, removing mutants that are likely to be
irrelevant to developers, and limiting the number of mutants per line
and per code review process; (3) Mutants are selected based on the historical
performance of mutation operators, further eliminating irrelevant mutants
and improving mutant quality.
This paper empirically validates the proposed approach by analyzing its
effectiveness in a code-review-based
setting, used by more than 24,000 developers on more than 1,000 projects.
%
The results show that the proposed approach produces
orders of magnitude fewer mutants and that context-based mutant filtering
and selection improve mutant quality and actionability.
Overall, the proposed approach represents a mutation testing framework that
seamlessly integrates into the software development workflow and is
applicable up to large-scale industrial settings.

\end{abstract}

}

\maketitle

\IEEEdisplaynontitleabstractindextext

%
\IEEEpeerreviewmaketitle

\section{Introduction}

Software testing is the predominant technique for ensuring software quality, and
various approaches exist for assessing test suite efficacy (i.e., a test suite's
ability to detect software defects). One such approach is code coverage, which is
widely used at Google~\cite{IvankovicPJF2019} and measures the degree to which a
test suite exercises a program. Code coverage is intuitive, cheap
to compute, and well supported by commercial-grade tools. However, code coverage alone might be misleading, in particular
when program statements are covered but the expected program outcome is not asserted
upon~\cite{offutt1996subsumption,checkedcoverage}.
Another approach is \defTerm{mutation analysis}, which systematically
seeds artificial faults into a program and measures a test suite's ability to
detect these artificial faults, called \defTerm{mutants}~\cite{demillo1978hints}.
Mutation analysis addresses the limitations of code coverage
and is widely considered the best approach for evaluating test suite
efficacy~\cite{andrews2006using,just2014mutants,ChenGTEHFAJ2020}.
\defTerm{Mutation testing} is an iterative testing approach that builds on top of mutation
analysis and uses undetected mutants as concrete test goals for which to create test
cases.

As a concrete example, consider the following fully covered, yet weekly tested, function:
\begin{lstlisting}[style=diffs, framesep=4pt, xleftmargin=4pt,escapechar=@]
public Buffer view() {
  Buffer buf = new Buffer();
  @\color{red}{buf.Append(this.internal\_buf); //mutation: delete this line}@
  return buf;
}
\end{lstlisting}
The tests only exercise the function, but do not assert upon its effects on
the returned buffer. This is just one example where mutation testing
outperforms code coverage: even though the line that appends some content
to \<buf> is covered, a developer is not informed about the fact that no test
checks for its effects. The statement-deletion mutation,
on the other hand, explicitly points out this testing weakness.

Google always strives to improve test quality, and thus decided to implement and deploy a mutation system to evaluate its effectiveness.
The sheer scale of Google's monolithic repository with approximately 2 billion lines of code~\cite{45424}, however, rendered the traditional approach to mutation testing infeasible: More than 500,000,000 test executions per day are gatekeepers for 60,000 change submissions to this code base,
ensuring that 13,000 continuous integrations remain healthy on a daily basis~\cite{swampUP}.
%
First, at this scale, systematically mutating the entire code base would create
far too many mutants, each potentially requiring many tests to be executed.
Second, neither the traditionally computed mutant-detection ratio, which quantifies test suite efficacy, nor simply showing all mutants that have evaded detection to a developer would be actionable.
Given that evaluating and resolving a single mutant takes several
minutes~\cite{SchulerZellerUncoveringICST2010,PetrovicIKAJ2018}, the required
developer effort for resolving all undetected mutants would be prohibitively expensive.

To make matters worse, even when applying sampling techniques to substantially reduce the
number of mutants, developers at Google initially classified \unprodStartPhaseOne of reported
mutants as unproductive. An \defTerm{unproductive} mutant is either trivially
equivalent to the original program or it is detectable, but
adding a test for it would not improve the test suite~\cite{PetrovicIKAJ2018}.
For example, mutating the initial
capacity of a Java collection
(e.g., \verb|new ArrayList(64)| $\mapsto$ \verb|new ArrayList(16)|) creates an unproductive
mutant. While it is possible to write a test that asserts on the
collection capacity or expected memory allocations, it is unproductive to do so.
In fact, it is conceivable that these tests, if written and added, would even have
a negative impact because their change-detector nature (specifically testing the current
implementation rather than the specification) violates testing best practices
and causes brittle tests and false alarms. 

Faced with the two major challenges in deploying mutation testing---the
computational costs of mutation analysis and the fact that
\emph{most mutants are unproductive}---we have developed a mutation testing approach that is scalable and usable, based on three central ideas:
\begin{compactenum}
	\item Our approach \emph{performs mutation testing on code changes}, considering only \emph{changed lines} of code (Section~\ref{mutation_testing}, based on our prior work~\cite{state_of_mt_at_google}), and surfacing mutants during code review. This greatly reduces the number of lines in
which mutants are created and matches a developer's unit of work for which additional tests are desirable.
	%
	\item Our approach uses \emph{transitive mutant suppression}, 
          using heuristics based on developer feedback (Section~\ref{arid}, based on our prior work~\cite{state_of_mt_at_google}). 
		  %
		  The feedback of more than 20,000 developers on thousands of
		  mutants over siz years enabled us to develop heuristics for mutant suppression
		  that improved the ratio of productive mutants from \prodStartPhaseOne to \prodEndPhaseTwo.
	%
	\item Our approach uses \emph{probabilistic, targeted mutant selection}, surfacing a restricted number of mutants based on historical performance (Section~\ref{sec:criteria}), further avoiding unproductive mutants.
\end{compactenum}

Based on an evaluation of the proposed mutation testing framework
on almost \ballparkMutantsGenerated mutants and \ballparkCLsTotal changes,
which surfaced \ballparkMutantsSurfaced mutants during code
review (Section~\ref{eval}), we conclude that, taken together, these
improvements make mutation testing feasible---even for
industry-scale software development environments.
%


%

%
%

\begin{figure*}[t!]
  \includegraphics[width=\textwidth]{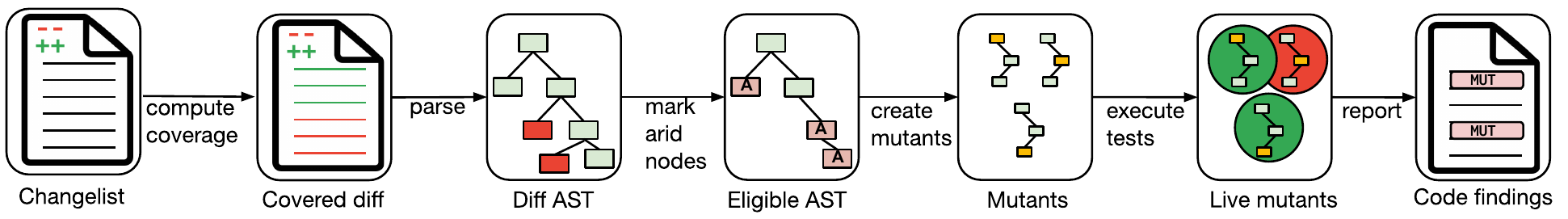}
  \caption{Mutagenesis process: (1) For a given changelist, line coverage is computed and code is parsed into an AST. (2) For AST nodes spanning covered lines, arid nodes are tagged as unproductive using the arid node detection heuristic. (3) Non-arid (eligible) nodes are mutated and tested. (4) Surviving mutants are surfaced as code findings.}
  \label{fig:mutagenesis}
\end{figure*}

\section{Mutation Testing at Google}
\label{mutation_testing}

Mutation testing at Google faces challenges of scale, both in terms of
computation time as well as integration into the developer workflow. Even
though existing work on selective mutation and other optimizations can
substantially reduce the number of mutants that need to be analyzed, it remains
infeasibly expensive to compute the absolute mutation score for the codebase at
any given fixed point due to the size of the code repository. It would be even
more expensive to keep re-computing the mutation score in any fixed time period
(e.g., daily or weekly), and it is impossible to compute the full score after
each commit. In addition to the computational costs of the mutation score, we
were also unable to find a good way to surface it to the developers in an
actionable way, as it is neither concrete nor actionable, and it does not guide testing. The scale, however, also makes surfacing individual mutants to developers challenging, in particular in light of unproductive mutants.
Mutation testing at Google is designed to overcome these challenges of scale and unproductive mutants, and therefore differs from the traditional approach to
mutation testing, described in the literature~\cite{offutt2001mutation}. 

Figure~\ref{fig:mutagenesis} summarizes how the Mutation
Testing Service at Google creates and analyzes mutants: Mutation testing is started when developers send changelists for code review.
A changelist is an atomic update to the version control system, and it consists of a list of files, the operations to be
performed on these files, and possibly the file contents to be modified or added, along with metadata like change description, author, etc.
First, the \MTS calculates the code
coverage for the changelist (Section~\ref{sec:coverage}). Then, it creates mutants (Section~\ref{sec:mutagenesis}) by 
 determining which nodes of the abstract syntax tree (AST) are eligible for
mutation. An AST node is eligible for mutation if it is covered by at least one
test and if it is not \emph{arid} (i.e., if mutated, it does not create
unproductive mutants; see Section~\ref{arid}).
The service then generates, executes, and analyzes mutants for all eligible AST nodes (Section~\ref{sec:analysis}). 
In the end, only a restricted set of surviving mutants is selected to be surfaced to the developer as part of the code review process (Section~\ref{codereview}).
This section describes the overall infrastructure and workflow of
mutation testing at Google.


%



\subsection{Prerequisites: Changelists and Coverage}\label{sec:coverage}

To enable mutation testing at Google, we implemented \emph{diff-based}
mutation testing: Mutants are only generated for lines that are changed. 
Once a developer is happy with their changelist, they send it to peers
for code review. At this point, various static and dynamic analyses are run for that
changelist and report back useful findings to the developer and the reviewers.
Line coverage is one such analysis: During code reviews, \emph{overall} and
\emph{delta} code coverage is surfaced to the
developers~\cite{IvankovicPJF2019}. Overall code coverage is the ratio of the
number of lines covered by tests in the file to the total number of
instrumented lines in the file. The number of instrumented lines is usually
smaller than the total number of lines, since artifacts like comments or pure
whitespace lines are not applicable for testing. Delta coverage is the ratio of
the number of lines covered by tests in the added or modified lines in the
changelist to the total number of added or modified lines in the changelist.

\begin{figure}
  \centering
  \includegraphics[width=\columnwidth]{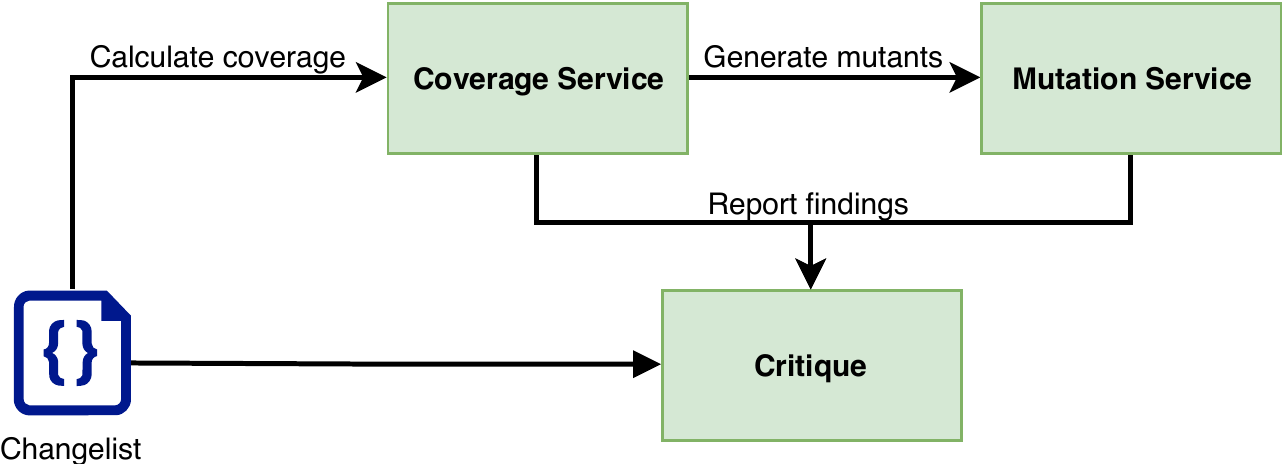}
  \caption{Code coverage and mutation testing integration}
  \label{fig:covmuta}
\end{figure}

Code coverage is a prerequisite for running mutation analysis, as shown in
Figure~\ref{fig:covmuta}, because of the high cost of generating and evaluating
mutants in uncovered lines, all of which would inevitably survive because the
code is not tested. Once line-level coverage is available for a changelist,
mutagenesis is triggered.

Google uses Bazel as its build system~\cite{bazel}. Build targets list their
sources and dependencies explicitly. Test targets can contain multiple tests,
and each test suite can contain multiple test targets. Tests are executed in
parallel.
Using the explicit dependency and source listing, test coverage analysis
provides information about which test target covers which line in the source
code. Results of coverage analysis link lines of code to a set of tests
covering them. Line level coverage is used during the test execution phase, where it
determines the minimal set of tests that need to be run in an attempt to kill a mutant.

\subsection{Mutagenesis}
\label{sec:mutagenesis}

Once delta coverage and line-level coverage metadata is available, the system
generates mutants in affected covered lines. Affected lines are added or
modified lines in the changelist, and covered lines are defined by the coverage
analysis results. 
The mutagenesis service receives a
request to generate \emph{point mutations}, i.e., mutations that produce a
mutant which differs from the original in one AST node on the requested line.
For each programming language supported, a special mutagenesis service capable
of navigating the AST of a compilation unit in that language accepts point
mutation requests and replies with potential mutants. 

For each point mutation request,
i.e., a $(file, line)$ tuple, a mutation operator is selected and a mutant is generated in that line if that mutation operator
is applicable to it. If no mutant is generated by the mutation operator, another is selected and so on until either a mutant is generated or all
mutation operators have been tried and no mutant could be generated. There are two mutation operator selection strategies, \textbf{random} and \textbf{targeted},
described in Section~\ref{sec:criteria}.

When a mutagenesis service receives a point mutation request, it first
constructs an AST of the file in question, and visits each node, labeling arid
nodes (Section~\ref{arid}) in advance using heuristics accumulated using
developer feedback about mutant productivity over the years.
Arid nodes are not considered for mutation
and no mutants are produced in them. Arid node labeling happens before
mutagenesis is started; mutants in arid nodes are not generated and discarded,
they are never created in the first place.

The \MTS implements mutagenesis for \numPLs programming
languages: \listPLs. For
each language, the service implements five mutation operators:
AOR (Arithmetic operator replacement),
LCR (Logical connector replacement),
ROR (Relational operator replacement),
UOI (Unary operator insertion), and
SBR (Statement block removal).
These mutation operators were originally introduced for Mothra~\cite{offutt1996experimental},
and Table~\ref{fig:tab_mut} gives further details for each. In Python, the unary
increment and decrement are replaced by a binary operator to achieve the same
effect due to the language design.
In our experience, the ABS (Absolute value insertion) mutation operator was reported
to predominantly create unproductive mutants, mostly because it acted on time-and-count related
expressions that are positive and nonsensical if negated, and is therefore not
used. Note that this is due to the style and features of our codebase, and may
not hold in general.

For each file in the changelist, a set of mutants is
requested, one for each affected covered line. Mutagenesis is performed
by traversing the ASTs in each of the languages, and
decisions are often done on the AST node level because it allows for
fine-grained decisions due to the amount of context available.

\begin{table*}[h]
\centering
\caption{Mutation operators implemented in the Mutation Testing Service}
\label{fig:tab_mut}
\begin{tabular}{lllcl}
\toprule
& \textsc{Name} & \multicolumn{3}{l}{\textsc{Scope}} \\
\midrule
  AOR & Arithmetic operator replacement & \code{\<a + b>}            & $\rightarrow$ & $\left\{\code{a, b, a - b, a * b, a / b, a \% b}\right\}$ \\
  LCR & Logical connector replacement   & \code{\<a \&\& b>}         & $\rightarrow$ & $\left\{\code{a, b, a || b, true, false}\right\}$ \\
  ROR & Relational operator replacement & \code{\<a \textgreater~b>} & $\rightarrow$ & $\left\{\code{a < b, a <= b, a >= b, true, false}\right\}$ \\
  UOI & Unary operator insertion        & \code{\<a>}                & $\rightarrow$ & $\left\{\code{a++, a-{}-}\right\}$; \code{b $\rightarrow$ !b} \\
  SBR & Statement block removal         & \code{\<stmt>}             & $\rightarrow$ & $\emptyset$ \\
\bottomrule
\end{tabular}
\end{table*}

\subsection{Mutation Analysis and Selection}

\label{sec:analysis}

Once all mutants are generated for a changelist, a temporary state of the
version control system is prepared for each of them, based on the original
changelist, and then tests are executed in parallel for all those states. This
makes for an efficient interaction and caching between our version control
system and build system, and evaluates mutants in the fastest possible manner.
Once test results are available, we randomly pick mutants from all surviving
mutants to be reported. We limit the number of reported mutants to at most 7
times the number of total files in the changelist, to ensure that the cognitive
overhead of understanding the reported mutants is not too high, which might
cause developers to stop using mutation testing. 7 is a result of
heuristics collected over the years of running the system. Selected surviving mutants are
reported in the code review UI to the author and the reviewers.

\subsection{Surfacing Mutants in the Code Review Process}
\label{codereview}

The selected mutants are shown to developers during the code review process. Most
changes to Google's monolithic codebase, except for a limited number of fully
automated changes, are reviewed by developers before they are merged into the
source tree. Potvin and Levenberg~\cite{45424} provide a comprehensive overview
of Google's development ecosystem.
Reviewers can leave comments on the changed code that must be resolved by the
author. A special type of comment generated by an automated analyzer is known
as a \emph{finding}. Unlike human-generated comments, findings do not need to
be resolved by the author before submission, unless a human reviewer marks them
as mandatory. Many analyzers are run automatically when a changelist is sent for
review: linters, formatters, static code and build dependency analyzers etc.
The majority of analyzers are based on the Tricorder code analysis
platform~\cite{43322}.  We display mutation analysis results during the code
review process because this maximizes the probability that the results will be
considered by the developers.

\begin{figure}
  \centering
  \includegraphics[width=0.8\columnwidth]{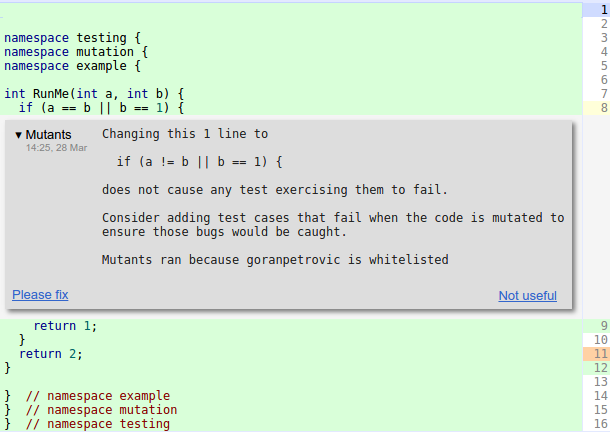}
  \caption{Mutant shown in the code review tool}
  \label{fig:critique}
\end{figure}

The number of comments displayed during code review can be large, so it is
important that all tools only produce high quality findings that can be used
immediately by the developers.  Surfacing non-actionable findings during code
review has a negative impact on the author and the reviewers. If an automated
changelist analyzer finding (e.g., a surviving mutant) is not perceived as useful,
developers can report that with a single click on the finding. If any of the
reviewers consider a finding to be important, they can indicate that to the
changelist author with a single click. Figure~\ref{fig:critique} shows an example
mutant displayed in Critique, including the ``Please Fix'' and ``Not useful''
links in the bottom corners. This feedback is accessible to the owner of the
system that created the findings, so quality metrics can be tracked and unhelpful
findings triaged, and ideally prevented in the future.

\subsection{Mutation Testing in Use at Google}

\begin{figure}[tb!]
  \centering
  \includegraphics[width=\columnwidth]{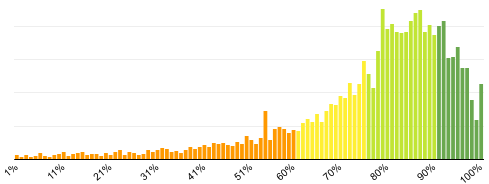}
  \caption{Distribution of project statement coverage}
  \label{fig:coverage_distribution}
\end{figure}

Google has a large codebase with code in various programming languages. The coverage distribution per project is
shown in Figure~\ref{fig:coverage_distribution}. Although the statement
coverage of most projects is satisfactory, even with our system that does heavy suppression and selection,
the number of live mutants per changelist is still significant (median is 2 mutants, 99\textsuperscript{th} percentile is 43
mutants).
To be of any use to the author and the reviewers, code findings need to be
surfaced quickly, before the review is complete. To further reduce the number
of mutants, mutations are never generated in uninteresting, arid lines, as
described in Section~\ref{arid}; furthermore, we probabilistically select mutants based on their historical mutation operator performance (Section~\ref{sec:criteria}).

\section{Arid Node Detection}
\label{arid}

Some parts of the code are less interesting than others. Surfacing live mutants
in uninteresting statements, for example debug logging statements, has a
negative impact on human time spent analyzing the finding, and its cognitive
overhead.  Because developers do not perceive adding tests to kill mutants in
uninteresting nodes as improving the overall efficacy of the suite to detect
faults, such mutants tend to survive. This section proposes an approach for
mutant suppression and a set of heuristics for detecting AST nodes in which
mutation is to be suppressed. There is a trade-off between correctness and
usability of the results; the proposed heuristic may suppress mutation in
relevant nodes as a side-effect of reducing uninteresting node mutations. We
argue that this is a good trade-off because the number of possible mutants is
always orders of magnitude larger than what we could reasonably present to the
developers within the existing developer tools, and it is more effective to
prevent high impact faults, rather than arid faults.

\subsection{Detecting Arid Nodes}

Mutation operators create mutants based on the AST of a program. The AST
contains nodes, which are statements, expressions or declarations, and their
child-parent relationships reflect their connections in the source
code~\cite{muchnick1997advanced}. In order to prevent the generation of unproductive mutants, we identify nodes in the AST that are related to uninteresting statements, i.e., arid nodes.

Most compilers differentiate simple and compound nodes in an AST. Simple nodes
have no body, e.g., a call expression names a function and parameters, but has no body. Compound nodes have at least one
body, e.g., a \texttt{for} loop might have a body, while an \texttt{if}
statement might have two: \texttt{then} and \texttt{else} branches. Examples of
arid nodes would be log statements, calls to memory-reserving functions like
\texttt{std::vector::reserve}, or writes to \texttt{stdout}; these
scenarios are typically not tested by unit tests.

The heuristic approach for labeling nodes as arid is two-fold and is defined in
Equation~\ref{eq:arid}:

\begin{equation}
  \emph{arid}(N) = \left\{\begin{matrix}
    \emph{expert}(N) & if$ $\emph{simple}(N)\\
    1\ if \bigwedge(\emph{arid}(b)) = 1, \forall b \in N & otherwise \\
\end{matrix}
\right.
\label{eq:arid}
\end{equation}

Here, $N \in T$ is a node in the abstract syntax tree $T$ of a program,
\emph{simple} is a boolean function determining whether a node is simple
(compound nodes contain their children nodes), and \emph{expert} is a boolean
function over a subset of simple statements in $T$ encoding manually curated
knowledge on arid simple nodes.
The first part of Equation~\ref{eq:arid} operates on simple nodes, is
represented by an expert curated manually for each programming language and is
adjusted over time. The second part operates on compound nodes and is defined
recursively. A compound node is an arid node iff \emph{all} of its parts
are arid.


The \emph{expert} function that flags simple nodes as arid is developed over time to
incorporate developer feedback on reported `Not useful' mutants. This process is
manual: if we decide a certain mutation is not productive and that the whole class
of mutants should not be created, the rule is added to the \emph{expert} function.
This is the critical part of the system because, without it, users would become
frustrated with non-actionable feedback and opt out of the system altogether.
Targeted mutation and careful surfacing of findings has been critical for adoption
of mutation testing at Google.
There are more than a hundred rules for arid node detection in our system.

\subsection{Expert Heuristic Categories}

\label{arid_example} The \emph{expert} function consists of various rules,
some of which are mutation-operator-specific, and some of which are
universal. We distinguish between heuristics that prevent the generation
of uncompilable vs. compilable yet unproductive mutants. Most heuristics deal
with the latter category, but the former is also important, especially in Go,
where the compiler is very sensitive to mutations (e.g., unused import is a
compiler error).
For compilable mutants, we distinguish between heuristics for
equivalent mutants, killable mutants, and redundant mutants, as
reported in Table~\ref{table_arid_heuristic}.

\subsubsection{Heuristics to Prevent Uncompilable Mutants}
A mutant should be a syntactically valid
program---otherwise, it would be detected by the compiler and not add any value
for testing. There are certain mutations, especially the ones that delete code,
that violate this validity principle. A prime example is deleting code in Go;
any unused variables or imported modules produce compiler errors. The proposed
heuristic is to gather all used symbols and put them in a slice instead of
deleting them so they are referenced and the compiler is appeased.

\subsubsection{Heuristics to Prevent Equivalent Mutants} 

Equivalent mutants, which are semantically equivalent to the mutated program,
are a plague in mutation testing and cannot generally be
detected automatically. However, there are some categories of equivalent
mutants that can be accurately detected. For example, in Java, the
specification for the \<size> method of a \texttt{java.util.Collection} is that
it returns a non-negative value. This means that mutations such as
\<collection.size() == 0> $\mapsto$ \<collection.size() <= 0> are guaranteed to produce an
equivalent mutant.

%

Another example for this category is related to memoization. Memoization
is often used to speed up execution, but its removal inevitably causes the
generation of equivalent mutants.
The following heuristic is
used to detect memoization: An \texttt{if} statement is a cache lookup if it is of the
form \code{if a, ok := x[v]; ok {return a}}, i.e., if a lookup in the map finds
an element, the \texttt{if} block returns that element (among other values, e.g.,
\texttt{Error} in Go). Such an \texttt{if} statement is a cache lookup statement and is
considered arid by the \emph{expert} function, as is its full body. The following example
shows a cache lookup in Go:

\begin{lstlisting}[style=diffs]
var cache map[string]string
func get(key string) string {
  @if val, ok := cache[key]; ok {
    return val
  }@
  value := expensiveCalculation(key)
  cache[key] = value
  return value
}
\end{lstlisting}
Removing the \texttt{if} statement just removes caching, but does not change
functional behavior, and hence yields an equivalent mutant. The program still
produces the same output for the same input---albeit slower. Functional tests
are not expected to detect such changes.

As a third example, a heuristic in this category avoids mutations of time
specifications because unit tests rarely test for time, and if they
do, they tend to use fake clocks. Statements invoking sleep-like functionality,
setting deadlines, or waiting for services to become ready (like
gRPC~\cite{grpc} server's \texttt{Wait} function that is always invoked in RPC
servers, which are abundant in Google's code base) are considered arid by the
\emph{expert} function.
\begin{lstlisting}[language=C,style=diffs]
  sleep(H100H); rpc.set_deadline(H10H);
\end{lstlisting}
  \hrule
\begin{lstlisting}[language=C,style=diffs]
  sleep(@200@); rpc.set_deadline(@20@);
\end{lstlisting}

\subsubsection{Heuristics to Prevent Unproductive Killable Mutants} 

Not all code is equally important. Much of it can be mutated, and those
mutants could actually be killed, but such tests are not considered valuable
and will not be written by experienced developers; such mutants are bad test
goals. Examples of this category are increments of values in monitoring system
frameworks, low level APIs like \texttt{mkdir} or flag changes: these are easy
to mutate, easy to test for, and yet mostly undesirable as tests.

A common way to implement heuristics in this category is to match function
names; indeed we suppress mutants in calls to hundreds of functions, which is
responsible for the highest number of suppressions. The star example of this
category is a heuristic that marks any function call arid if the function name
starts with the prefix \texttt{log} or the object on which the function is
invoked is called \texttt{logger}. We validated this heuristic by randomly
sampling 100 nodes that were marked arid by the \texttt{log} heuristic, and
found that 99 indeed were correctly marked, while one had marginal utility. We
have fuzzy name suppression rules for more than 200 function families.


\begin{lstlisting}[language=go,style=diffs]
log.infof("network speed: %v", bytesH/Htime)
\end{lstlisting}
\hrule
\begin{lstlisting}[language=go,style=diffs]
log.infof("network speed: %v", bytes@+@time)
\end{lstlisting}

\subsubsection{Heuristics to Prevent Redundant Mutants} 

There has been a lot of research on redundant
mutants, targeted at reducing the cost of mutation testing.
While the cost aspect is not a concern for us, because we generate at most a single
mutant in a line, user experience and consistency are important concerns.
In a
code review context, we surface mutants in each snapshot; when the developers
update their code, possibly writing tests to kill mutants, we rerun mutation
testing on the new code and report new mutants.
Because of this, we suppress
some redundant mutants so that mutants are consistently reported, as opposed to
alternating between redundant mutants, which introduces cognitive overhead and can be
confusing. 


As an example, in C++, the LCR mutation operator has a special case when dealing
with \texttt{NULL} (i.e., \texttt{nullptr}), because of its logical equivalence with
\texttt{false}:
\begin{center}
\begin{tabular}{lcl}
\toprule
\textsc{Original node} & &\textsc{Potential mutants} \\
\midrule
\multirow{5}{*}{\code{if (x != nullptr)}} &
\multirow{5}{*}{$\longmapsto$}
& \<if (x)> \\
&& \textbf{\<if (nullptr)>} \\
&& \<if (x == nullptr)> \\
&& \textbf{\<if (false)>} \\
&& \<if (true)> \\
\bottomrule
\end{tabular}
\end{center}
The mutants marked
in bold are redundant (equivalent to one another) because the value of \<nullptr> is equivalent to
\<false>. Likewise, the opposite example, where the condition is \texttt{if (nullptr ==
x)}, yields redundant mutants for the left-hand side.
These mutations are suppressed.

\subsubsection{Experience with Heuristics}
\label{sec:experience_heuristics}
The highest mutant productivity gains came from the three heuristics implemented
in the early days: suppression of mutations in
logging statements, time-related operations (e.g., setting deadlines, timeouts,
exponential backoff specifications etc.), and finally configuration flags. Most
of the early feedback was about unproductive mutants in such code, which is ubiquitous
in the code base. While it is hard to measure exactly, there is strong indication
that these suppressions account for improvements in productivity from about
\prodStartPhaseOne to \prodEndPhaseOne. Additional heuristics and refinements
progressivley improved producitvity to \prodEndPhaseTwo.

Heuristics are implemented by matching AST nodes with the full compiler
information available to the mutation operator. Some  heuristics are unsound: they employ
fuzzy name matching and recognize AST shapes, but can suppress a productive
mutant. On the other hand, some heuristics make use of the full type information
(like matching \texttt{java.util.HashMap::size} calls) and are sound. Sound
heuristics are demonstrably correct, but we have had much more important
improvements of perceived mutant usefulness from unsound heuristics.

\begin{table}[tb]
\centering
\caption{Arid node heuristics.}
\label{table_arid_heuristic}
\begin{tabular}{lll}
\toprule
  \textsc{Heuristic} & \textsc{Count} & \textsc{Frequency} \\
\midrule
  Uncompilable & 1 & Common \\
  Equivalent & 13  & Common \\
  Unproductive killable & 16 & Very common \\
  Redundant & 2 & Uncommon \\
\bottomrule
\end{tabular}
\end{table}

For a detailed list of heuristics, please refer to
Appendix~\ref{appendix:heuristics}.


\section{Mutant Selection Criteria}
\label{sec:criteria}

\begin{figure}
  \includegraphics[width=\columnwidth]{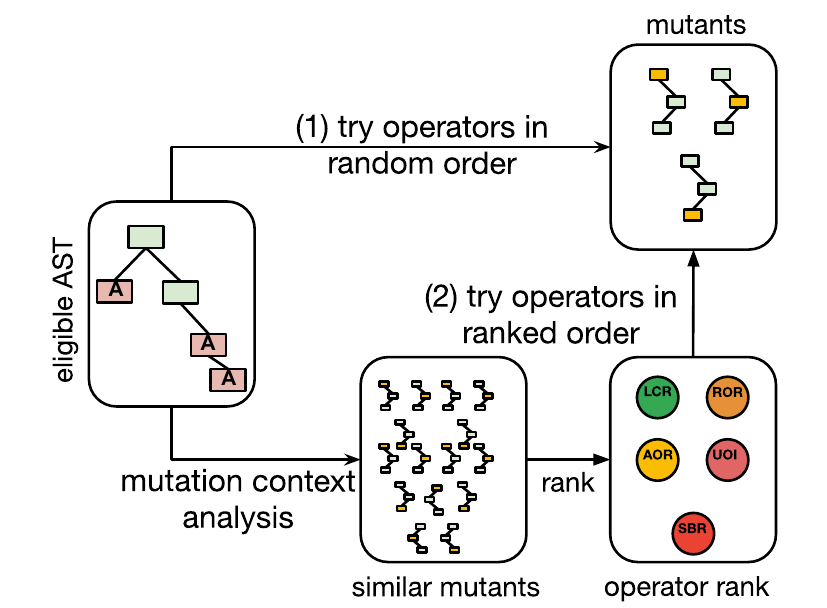}
  \caption{Random (1) vs. Targeted (2) mutation selection}
  \label{fig:targeted-mutation}
\end{figure}

Once arid nodes have been identified in the AST, the next step (cf.
Section~\ref{sec:mutagenesis}) is to produce mutants for the remaining,
non-arid nodes. There are two issues arising from this: First, only mutants
that survive the tests can be shown to developers, whereas those that are
killed just use computational resources. Many mutants never survive the test
phase, and are not reported to the developer and reviewers during code review.
An iterative approach, where after the first round of tests further rounds of
mutagenesis could be run for lines in which mutants were killed, would use the
build and test systems inefficiently, and would take much longer because of
multiple rounds.
Second, not all surviving mutants are equally productive: Depending on the
context, certain mutation operators may produce better mutants than others.
Reporting all surviving mutants for a line would prolong the mutagenesis step
and increase test evaluation costs in a prohibitive manner.
Because of this, effective selection criteria not only constitute a
good trade-off, but are crucial in making mutation analysis results actionable
during code review. 
In this section, we present a basic random selection strategy that generates one mutant per covered
line and considers information about arid nodes, and a targeted selection, which considers the past
performance of mutation operators in similar context (Figure ~\ref{fig:targeted-mutation}).

\subsection{Random Selection}
\label{strategy:random}

The basic principle of a random line-based mutant selection is shown in
Listing~\ref{algo_random}: For each line in a changelist, one of the mutants that can
be generated for that line would be selected randomly, or alternatively a mutation target is picked randomly first and then a mutation operator is randomly selected.

\begin{figure}[h]
  \begin{lstlisting}[language=Python,style=algo,caption={Na\"{i}ve random selection},captionpos=b,label={algo_random}]
  function Mutagenesis(diff_ast)
    mutants <= -0
    for line in covered_lines(diff_ast)
      mutants <= uniform_random(all_mutants(line))
    endfor
    return mutants
  \end{lstlisting}
\end{figure}

Since our approach to mutation testing is based on the identification of arid
nodes which should not be mutated, the random selection algorithm we use is
described in Listing~\ref{algo_suppressed_random}. For each language, the
Mutation Testing Service implements mutation operators as AST visitors. The  mutation operators available for a language are randomly shuffled, and are used one by one to try and create a mutant in the given file and line, until one succeeds. We do this for each changed line in the changelist that is covered by tests. If any mutant can be created in a line, one will be created in that line, but which one will depend on the random shuffle and the AST itself (e.g., in a line without
relational operators, the ROR mutation operator will not produce a mutant, but SBR might, because most lines can be deleted). If the first mutation operator in the randomly shuffled order cannot produce a mutant in a given line, either because it is not applicable to it, or because the relevant AST nodes are labeled arid, the next mutation operator is invoked, until either a mutant is produced or there are no more mutation operators left.
This is done for each mutation request.

\begin{figure}[h]
  \begin{lstlisting}[language=Python,style=algo,caption={Random selection with suppression},captionpos=b,label={algo_suppressed_random}]
  function Mutagenesis(diff_ast)
    mutants <= -0
    productive_ast = remove_arid_nodes(diff_ast)
    ops = shuffle({UOI, ROR, SBR, LCR, AOR})
    for line in covered_lines(productive_ast)
      for op in ops
        if can_generate(op, line)
          mutants U= generate_mutant(op, line)
          break
    return mutants
  \end{lstlisting}
\end{figure}

It is important to note that many nodes are labeled as arid by our
heuristic (see Section~\ref{arid}), and are not considered for mutation at all.
Furthermore, only a single mutant in a line is ever produced, all others are
not considered. These design decisions proved to be the core of making mutation
testing feasible at very large scale.

\subsection{Targeted Selection}
\label{strategy:targeted}

The targeted mutation operator selection strategy orders the operators by their
perceived productivity in the mutation AST context, as shown in
Listing~\ref{algo_targeted}.

\begin{figure}[h]
  \begin{lstlisting}[style=algo,caption={Targeted selection with suppression},captionpos=b,label={algo_targeted}]
  function Mutagenesis(diff_ast)
    mutants <= -0
    productive_ast = remove_arid_nodes(diff_ast)
    ops = {UOI, ROR, SBR, LCR, AOR}
    for line in covered_lines(productive_ast)
      ops = order_by_historic_productivity(line, ops)
      for operator in ops
        if can_generate(operator, line)
          mutants U= generate_mutant(operator, line)
          break
    return mutants
  \end{lstlisting}
\end{figure}

The information about how productive mutating a particular AST node by a particular
mutation operator is, is based on historical information: First, we can
determine a mutation operator's \emph{survivability} (i.e., the fraction of
mutants produced by the operator in the past that were not killed by the
existing tests) in a particular context. Second, we can determine a mutant's
\emph{productivity} using developer feedback: Each reported mutant can be
flagged as productive or unproductive by the author of the changelist or any of
the reviewers of the changelist. We consider this a strong signal because it
comes from experienced professionals that understand the context of the
mutant.

Using this information, we can order the mutation operators by
survivability and perceived productivity, rather than using a random shuffle.
For each mutant, an AST context is kept, describing the environment of the AST
node that was mutated, along with the productivity feedback and whether the
mutant was killed or not. When the mutagenesis service receives a point
mutation request, for nodes for which the mutation is requested, it finds
similar nodes from the body of millions of previously evaluated mutants using
the AST context, and then looks into historical performance of those mutants in
two categories: developer feedback on productivity and mutant survivabiliy.
Mutation operators are ordered using this metric rather than uniformly shuffled,
and mutagenesis is attempted in that order, to maximize the probability that
the mutant will be productive, or at least survive to be reported in the code
review. For example, if we are mutating a binary expression within an
\texttt{if} condition, we will find mutants done in a similar AST context and
see how each mutation operator performed in them.

\subsection{Mutation Context}\label{section:context}

In order to apply historical information about mutation productivity and
effectiveness, we need to decide how similar candidate mutations are compared
to past mutations. We define a mutation to be similar if it happened in a
similar context, e.g., replacing a relational operator within an \texttt{if}
condition that is the first statement in the body of a \texttt{for} loop, as shown in Listing~\ref{code}.

As an efficient means to capture the similarity of the context of two
mutations, we use the hashing framework for tree-structured data introduced by
Tatikonda et al.~\cite{tatikonda2010hashing}, which maps an unordered tree into
a multiset of simple structures referred to as $pivots$. Each pivot captures
information about the relationship among the nodes of the tree (see
Section~\ref{transformation}).

Finding similar mutation contexts is then reduced to finding similar pivot
multisets. To identify similar pivot multisets, we produce a
MinHash~\cite{minhash} inspired fingerprint of the pivot multiset. Because the
distance in the fingerprint space correlates with the distance in the tree
space, we can find similar mutation contexts efficiently by finding similar
fingerprints of the node under mutation.

\subsection{Generating Pivots from ASTs}\label{transformation}

In order to capture the intricate relationship between nodes in the AST, we translate
the AST into a multiset of pivots. A pivot is a triplet of nodes from the AST that
encodes their relationship; for nodes $u$ and $v$, a pivot $p$ is tuple $(lca, u, v)$, where
$lca$ is the lowest common ancestor of nodes $u$ and $v$. The pivot represents a subtree of the AST. The set of
all pivots involving a particular node describes the tree from the point of
view of that node. In mutation testing, we are only interested in nodes that
are close to the node being mutated, so we constrain the set of pivots to
pivots containing nodes that are a certain distance from the node considered for
mutation.

\begin{figure}
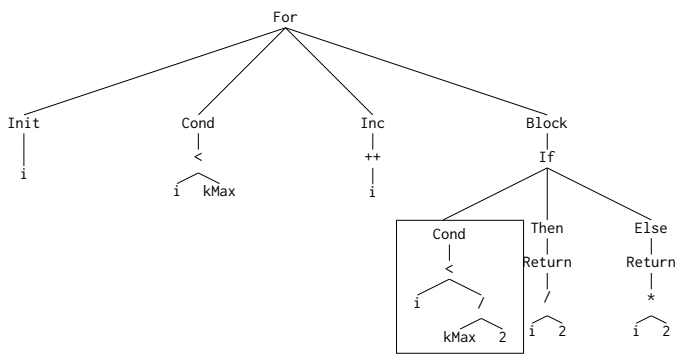

  \newcommand{\qlabelhook}{\small\texttt}
  \newcommand{\qleafhook}{\small\texttt}
  \resizebox{\linewidth}{!}{%
  \Tree [
    .For
      [.Init [i ]]
      [.Cond [.< i kMax ] ]
      [.Inc [.++ i ] ]
      [.Block [
        .If
          [.Cond [.< i [./ kMax 2 ] ] ] !{\qframesubtree}
          [.Then [.Return [./ i 2 ] ] ]
          [.Else [.Return [.* i 2 ] ] ]
        ]
      ]
    ]
  }
  \caption{AST for the C++ example in Listing \ref{code}}
  \label{fig:ast}
\end{figure}

In the example of replacing a relational operator in an \texttt{if} condition within a body
of the \texttt{for} loop in Listing~\ref{code}, one pivot might be $($\texttt{if}, \texttt{Cond}, $*)$, and another
$($\texttt{Cond}, $i$, \texttt{kMax}$)$. All combinations of two nodes within some distance from
the node being mutated in the AST in Figure~\ref{fig:ast} and their lowest common ancestor make pivot structures.

\begin{figure}[ht!]
  \begin{lstlisting}[language=Python,label={code},caption={C++ snippet with an \texttt{if} statement within a \texttt{for} loop}]
for (int i = 0; i < kMax; ++i) {
  if (i < kMax / 2) {
    return i / 2;
  } else {
    return i * 2;
  }
}
\end{lstlisting}
\end{figure}

Pivot multisets $P$ precisely preserve the structural relationship
of the tree nodes (\textit{parent-child} and \textit{ancestor} relations), so the tree similarity
of two AST subtrees $T1$ and $T2$ can be measured as the Jaccard index of the pivot multisets~\cite{tatikonda2010hashing} as shown in
equation~\ref{capcup}.

\begin{equation}
  \label{capcup}
  d(T1, T2) = \textrm{Jaccard}(P(T1), P(T2)) = \frac{|P(T1) \cap P(T2)|}{|P(T1) \cup P(T2)|}
\end{equation}

\subsection{Fingerprinting Pivot Multisets}

Pivot multisets are potentially quadratic in tree size, leading to costly union
and intersection operations. Even a trivial \texttt{if} statement with a single \texttt{return}
statement produces large pivot sets, and set
operations become prohibitive. To alleviate that, a fingerprinting function is
applied to convert large pivot multisets into fixed-sized fingerprints.

We hash the pivot sets to single objects that form the multiset of
representatives for the input AST. The size of the multiset can be large, especially
for large programs. In order to improve the efficiency of further manipulation,
we use a signature function that converts large pivot hash sets into shorter
signatures. The signatures are later used to compute the similarity
between the trees, taking into consideration only the AST node type and
ignoring everything else, like type data or names of the identifiers.

We use a simple hash function to hash a single pivot $p =
(lca, u, v)$ into a fixed-size value, proposed by Tatikonda and
Parthasarathy~\cite{tatikonda2010hashing}.

$$ h(p) = (a \cdot lca + b \cdot u + c \cdot v) \bmod K $$
$$ a, b, c \in \mathbb{Z_P} $$

For $a, b, c$ we pick small primes, and for $K$ a large prime that fits in 32
bits.  To be able to hash AST nodes, we assign sparse integer hash values to
different AST node types in each language, e.g., a C++ \texttt{FunctionDecl} is assigned
8500, and \texttt{CXXMethodDecl} 8600.  For nodes in the pivot $(lca, u, v)$ we use
these assigned hashes.

The signature for such a bag of representatives is generated using
a MinHashing technique. The set of pivots is permuted and hashed under that
permutation. To minimize the false positives and negatives (i.e., different trees hash to similar hashes, or vice versa),
this is repeated $k$ times, resulting in $k$-MinHashes.

The goal is that the signatures are similar for similar (multi)sets and
dissimilar for dissimilar ones. Jaccard similarity between two sets can be
estimated by comparing their MinHash signatures in the same way~\cite{minhash}, as shown in
equation~\ref{minhasheq}. The MinHash scheme can be
considered an instance of locality-sensitive hashing, in which ASTs that have a
small distance to each other are transformed into hashes that preserve that
property.

\begin{equation}
  \label{minhasheq}
  d(T1, T2) = \frac{|P(T1) \cap P(T2)|}{|P(T1) \cup P(T2)|} \approx \frac{|hash(T1) \cap hash(T2)|}{|hash(T1) \cup hash(T2)|}
\end{equation}

When mutating a node, we calculate its pivot set and hash it. We find similar AST contexts
using nearest neighbor search algorithms. We observe how different mutants
behave in this context and which mutation operators produce the most productive and surviving mutants.
This is the basis for targeted mutation selection.

\section{Evaluation}
\label{eval}

In order to bring value to developers, the \MTS at Google
needs to surface few productive mutants, selected from a large pool of
mutants---most of which are unproductive.
Recall that a productive mutant elicits an effective test, or otherwise advances
code quality~\cite{PetrovicIKAJ2018}.
Therefore, our goal is two-fold. First, we aim to select mutants with a high
survival rate and productivity to maximize their utility as test objectives.
Second, we aim to surface very few mutants to reduce computational effort and
avoid overwhelming developers with too many findings.

In addition to the design decision of applying mutation testing at the level of
\cls rather than projects, two technical solutions reduce the
number of mutants: (1) mutant suppression using arid nodes and (2)
one-per-line mutant selection. The first research question aims to answer how effective
these two solutions are:
\begin{itemize}
\item {\bf RQ1 Mutant suppression}. How effective is mutant suppression using
arid nodes and 1-per-line mutant selection? (Section~\ref{results_suppression})
\end{itemize}
To understand the influence of mutation operator selection on mutant survivability and productivity in the remaining non-arid nodes, we consider historical data, including developer feedback. We aim to answer the following two research questions:
\begin{itemize}
\item {\bf RQ2 Mutant survivability}. Does mutation operator selection influence the probability that a generated mutant survives the test suite? (Section~\ref{results_survival})
\item {\bf RQ3 Mutant productivity}. Does mutation operator selection influence developer feedback on a generated mutant? (Section~\ref{results_feedback})
\end{itemize}
Having established the influence of individual mutation operators on survivability and productivity, the final question is whether mutation context can be used to improve both. Therefore, our final research question is as follows:
\begin{itemize}
\item {\bf RQ4 Mutation context}. Does context-based selection of mutation operators improve mutant survivability and productivity? (Section~\ref{results_context})
\end{itemize}

\subsection{Experiment Setup}

For our analyses, we established two datasets, one with data on all mutants, and
one containing additional data on mutation context for a subset of all mutants.

\smallskip
\noindent\textbf{Mutant dataset.} The mutant dataset contains
\numMutantsGenerated mutants across \numPLs programming languages:
\listPLs.
Table~\ref{table_mutants_lang} summarizes the mutant dataset and gives the
number and ratio of mutants per programming language, the average number of mutants per
\cl and the percentage of mutants that survive the test suite.
Table~\ref{table_mutants_type} breaks down the numbers by mutation operator.
%

\begin{table}[t]
\centering
\caption{Summary of the mutant dataset. (Note that \listPLsExcl are excluded
from our analyses because of insufficient data.)}
\label{table_mutants_lang}
\vspace*{-6pt}
\begin{tabular}{lrrrr}
\toprule
  \textsc{Language} & \multicolumn{3}{c}{\textsc{Generated mutants}} & \textsc{Survivability}   \\
\cmidrule{2-4}
  & \textsc{Count} & \textsc{Ratio} & \textsc{Per CL} & \\
\midrule
C++ & 7,197,069 & 42.5\% &  23.2 & 12.5\%\tabularnewline
Java & 2,894,772 & 17.1\% &  14.8 & 13.2\%\tabularnewline
Go & 1,988,798 & 11.7\% &  27.6 & 12.5\%\tabularnewline
Python & 1,689,382 & 10.0\% &  21.3 & 13.2\%\tabularnewline
TypeScript & 1,006,531 &  5.9\% &  20.8 & 10.8\%\tabularnewline
JavaScript &   908,014 &  5.4\% &  31.0 &  9.4\%\tabularnewline
Dart &   581,109 &  3.4\% &  17.4 & 16.3\%\tabularnewline
\midrule
\excl SQL &   478,975 &  2.8\% &  91.2 & 11.7\%\tabularnewline
\excl Common Lisp &   148,289 &  0.9\% & 179.3 &  2.2\%\tabularnewline
\excl Kotlin &    42,209 &  0.2\% &  20.7 & 11.0\%\tabularnewline
\midrule
Total & 16,935,148 & 100\% & 21.8 & 12.5\%\tabularnewline

\bottomrule
\end{tabular}
\end{table}

\begin{table}[t]
\centering
\caption{Number of mutants per mutation operator.}
\label{table_mutants_type}
\vspace*{-6pt}
\begin{tabular}{lrrrr}
\toprule
  \textsc{Operator} & \multicolumn{2}{c}{\textsc{Generated mutants}} & \textsc{Survivability}   \\
\cmidrule{2-4}
  & \textsc{Count} & \textsc{Ratio} \\
\midrule
SBR & 11,522,932 & 68.0\% & 12.7\%\tabularnewline
UOI &  3,137,375 & 18.5\% &  9.6\%\tabularnewline
LCR &  1,305,499 &  7.7\% & 16.3\%\tabularnewline
ROR &    672,009 &  4.0\% & 14.7\%\tabularnewline
AOR &    297,333 &  1.8\% & 13.5\%\tabularnewline
\midrule 
Total & 16,935,148 & 100\% & 12.5\%\tabularnewline

\bottomrule
\end{tabular}
\end{table}

We created this dataset by gathering data on all mutants that the \MTS
generated since its inauguration, which refers to the date when we made the
service broadly available, after the initial development of the service and its
suppression rules (see Section~\ref{sec:experience_heuristics}).
We did not perform any data
filtering, hence the dataset provides information about all mutation
analyses that were run.

In total, our data collection considered \numCLs \cls that were part of the code review
process. For these, \numMutantsGenerated mutants were generated, out of which
\numMutantsSurfaced were surfaced. Out of all
surfaced mutants, \numMutantsFeedback received explicit developer
feedback.
For each considered \cl, the mutant dataset contains information about:

\begin{itemize}
  \item affected files and affected lines,
  \item test targets testing those affected lines,
  \item mutants generated for each of the affected lines,
  \item test results for the file at the mutated line, and
  \item mutation operator and context for each mutant.
\end{itemize}

Our analysis aims to study the efficacy and perceived productivity
of mutants and mutation operators across programming languages. Note that our
mutant dataset is likely specific to Google's code style and review
practices. However, the code style is widely adopted~\cite{googlestyleguide},
and the modern code review process is used throughout the industry~\cite{bacchelli2013expectations}.

Information about mutant survivability per programming language or mutation
operator can be directly extracted from the dataset and allows us to answer
research questions \textbf{RQ1}, \textbf{RQ2} and \textbf{RQ3}.

\smallskip
\noindent\textbf{Context dataset.} The context dataset contains
\numMutantsContext mutants (a subset of the mutant dataset) for the
top-four programming languages: C++, Java, Go, and Python. Each mutant in this
dataset is enriched with the information of 
whether our context-based selection strategy would have selected that mutant.
When generating mutants, we would also run the context-based prediction, and we
persisted the prediction information along with the mutants. If the randomly chosen
operator was indeed what the prediction service picked, this mutant is the one
with the highest predicted value.
For each mutant, the dataset contains:

\begin{itemize}
  \item all information from the mutant dataset,
  \item predicted survivability and productivity for each mutation in similar context, and
  \item information about whether the mutant has the highest predicted survivability/productivity.
\end{itemize}

We created this dataset by using our context-based mutation selection strategy
during mutagenesis on all mutants during a limited period of time. During this
time, we automatically annotated the mutants, indicating whether a mutant would
be picked by the context-based mutation selection strategy along with the mutant
outcome in terms of survivability and productivity.
This dataset enables the evaluation of our context-based mutation selection
strategy and allows us to answer research question \textbf{RQ4}.

\smallskip
\noindent\textbf{Experiment measures:}
Surviving the initial test suite is a precondition for surfacing a mutant, but
survivability alone is not a good measure of mutant productivity. For example,
a mutation that changes the timeout of a network call likely survives the test
suite but is also very likely to be unproductive (i.e., a developer will not consider writing a
test for it). Hence, developer feedback indicating that a mutant is indeed
(un)productive is a stronger signal.

We measure mutant productivity via user feedback gathered from Critique
(Section~\ref{codereview}), where each
surfaced mutant displays a \textit{Please fix} (productive mutant) and a \textit{Not
useful} (unproductive mutant)
link. \textit{Please fix} corresponds to a request to the author of a \cl to improve the
test suite based on the surfaced mutant; \textit{not useful} corresponds to a false alarm
or generally a non-actionable code finding.
\percentMutantsPleaseFix of all surfaced mutants with feedback were labeled as
productive by developers.
Note that this ratio is an aggregate over the entire data set. Since the
inauguration of the \MTS, productivity
has increased over time from \prodStartPhaseTwo to
\prodEndPhaseTwo because we generalized the feedback
on unproductive mutants and created suppression rules for the $expert$
function, described in Section~\ref{arid}. This means that later mutations of
nodes in which mutants were found to be unproductive will be suppressed,
generating fewer unproductive mutants over time.
Surfaced mutants without explicit developer feedback are not considered
for the productivity analysis.

\subsection{RQ1 Mutant Suppression}
\label{results_suppression}

\begin{figure}[tb]
\centering
\includegraphics[trim=0 20 0 40,clip,width=.9\linewidth]{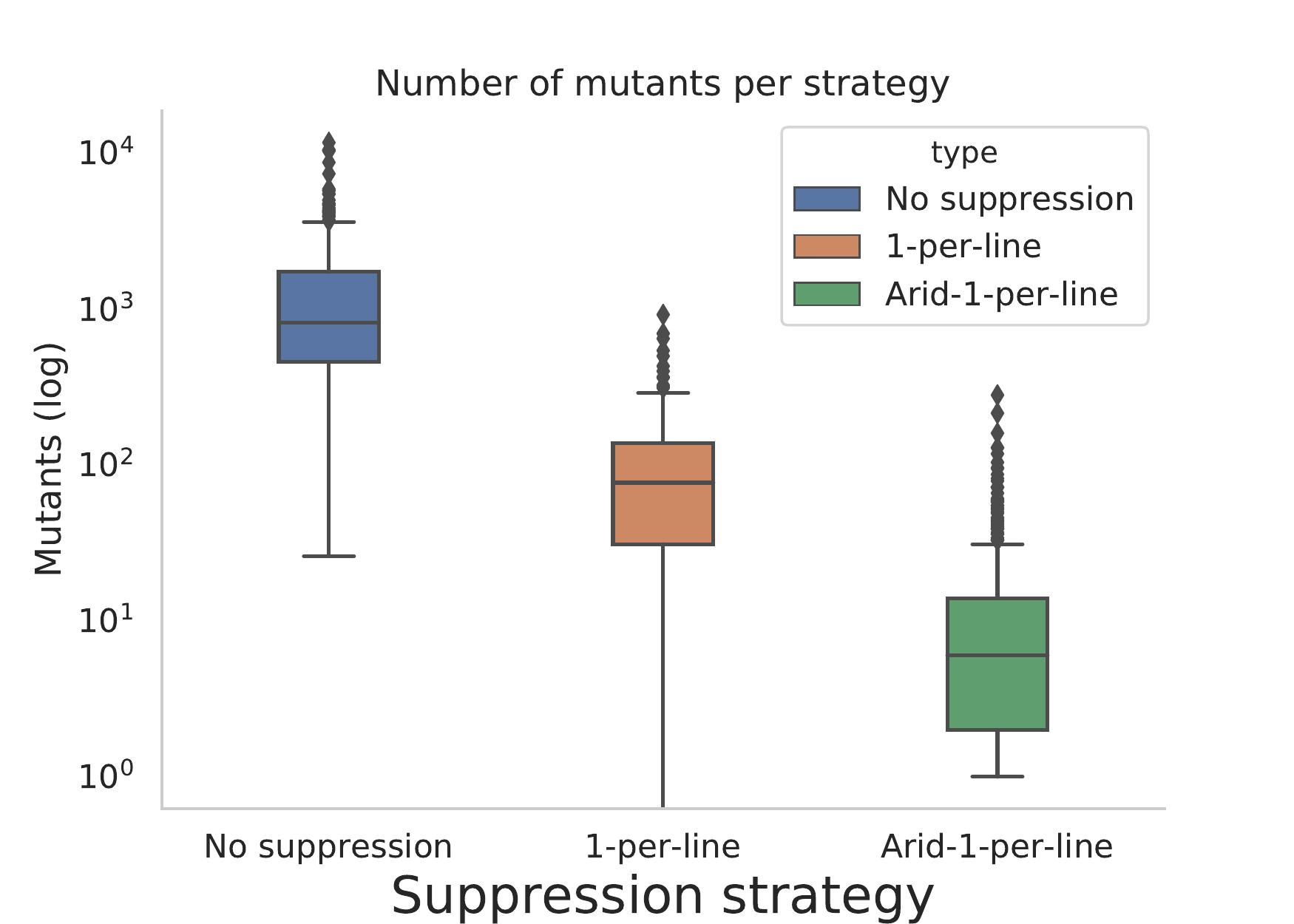}
\caption{Number of generated mutants per \cl for no suppression (traditional mutagenesis),
1-per-line and arid-1-per-line (our approach). (Note the log-scaled vertical
axis.)
  }
\label{fig:comparison}
\end{figure}

In order to compare our mutant-suppression approach with the traditional
mutagenesis, we (1) randomly sampled 5,000 \cls from the mutant
dataset, (2)
determined how many mutants traditional mutagenesis produces, and (3) compared
the result with the number of mutants generated by our approach. (Since
traditional mutation analysis is prohibitively expensive at scale, we adapted
our system to only generate all mutants for the selected \cls.)
Figure~\ref{fig:comparison} shows the results for three strategies: no
suppression (traditional), select one mutant per line, and select
one mutant per line after excluding arid nodes (our approach). We include the
1-per-line approach in the analysis to evaluate the individual contribution of
the arid-node suppression, beyond sampling one mutant per line.

As shown in and Table~\ref{tab:suppression}, the
median number of generated mutants is 820 for traditional mutagenesis, 77 for
1-per-line selection, and only 7 for arid-1-per-line selection. Hence, our
mutant-suppression approach reduces the number of mutants by two orders of
magnitude. Table~\ref{tab:suppression} also shows the results for a Mann-Whitney
U test, which confirms that the distributions are statistically significantly
different.

\begin{table}
 \caption{Mann-Whitney U test comparing the distributions of the number of mutants generated by different strategies.}
\label{tab:suppression}
	\resizebox{\columnwidth}{!}{%
\begin{tabular}{llrrr}
\toprule
  \textsc{Strategy A} & \textsc{Strategy B} & \textsc{p-value} & \textsc{Median A} & \textsc{Median B} \\
\midrule
No suppression & 1-per-line & <.0001 & 820 & 77 \\
1-per-line & Arid-1-per-line & <.0001 & 77 & 7 \\
No suppression & Arid-1-per-line & <.0001 & 820 & 7 \\
\bottomrule
\end{tabular}
}
\end{table}

Our mutant-suppression approach generates fewer than 20 mutants for most \cls;
the 25th and 75th percentiles are 3 and 19, respectively. In contrast, the 25th
and 75th percentiles for 1-per-line are 31 and 138 mutants. Traditional
mutagenesis generates more than 450 mutants for most \cls (the 25th and 75th
percentiles are 460 and 1734, respectively), further underscoring that this
approach is impractical, even at the \cl level. Presenting hundreds of mutants,
most of which are not actionable, to a developer would almost certainly result
in that developer abandoning mutation testing altogether.

\begin{result}
	\textbf{RQ1:} Arid-node suppression and 1-per-line selection significantly
    reduce the number of mutants per \cl, with a median of only 7 mutants per
    \cl (compared to 820 mutants for traditional mutagenesis).
\end{result}

\subsection{RQ2 Mutant Survivability}
\label{results_survival}

\begin{figure}[tb]
\centering
\subfloat[Survivability per programming language.]{
\includegraphics[trim=0 20 0 0,clip,width=.4\textwidth]{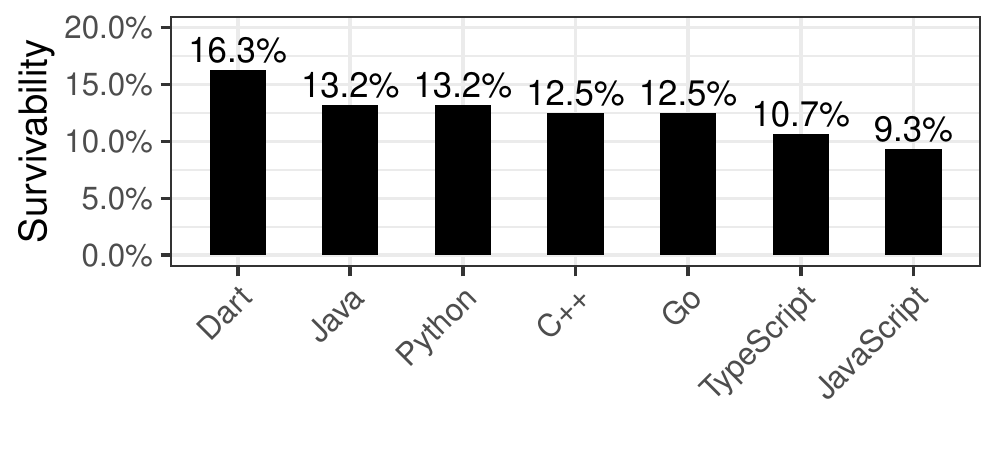}
}

\subfloat[Survivability per mutation operator.]{
\includegraphics[trim=0 20 0 0,clip,width=.4\textwidth]{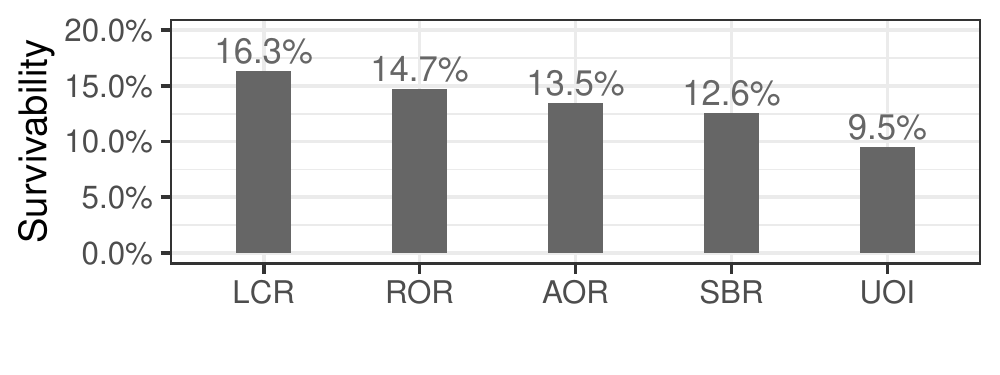}
}
\caption{Mutant survivability.}
\label{fig:barcharts}
\end{figure}

Mutant survivability is important because we generate at most a single
mutant per line---if that mutant is killed, no other mutant is generated.
To be actionable,
mutants have to be reported as soon as possible in the code review process, as described in
Section~\ref{sec:criteria}. Therefore, we aim to maximize mutant survivability
because it directly impacts the number of surfaced mutants.

Overall, \percentMutantsKilled of all generated mutants are killed by the initial test suite. Note that this is
not the same as the traditional mutation
score~\cite{DeMillo:1978:HTD:1300736.1301357} (ratio of killed mutants
to the total number of mutants) because mutagenesis is probabilistic and only
generates a subset of all mutants. This means only a
fraction of all possible mutants are generated and evaluated, and many
other mutants are never generated because they are associated with arid nodes.

Tables~\ref{table_mutants_lang} and~\ref{table_mutants_type} show the
distribution of number of mutants and mutant
survivability, broken down by programming language and mutation operator.
Figure~\ref{fig:barcharts} visualizes the mutant survivability data.
Because the SBR mutation operator can be applied to almost any
non-arid node in the code, it is no surprise that this mutation operator
dominates the number of mutants,
contributing roughly 68\% of all mutants. While SBR is a prolific and versatile
mutation operator, it is also the second least likely to survive the test suite:
when applicable to a \cl, SBR mutants are surfaced during code review with a
probability of 12.6\%.
Overall, mutant
survivability is similar across mutation operators, with a notable exception of
UOI, which has a survivability of only 9.5\%.
Mutant
survivability is also similar across programming languages with the exception
of Dart, whose mutant survivability is noticeably higher.
We conjecture that this is because Dart is mostly used for web development which
has its own testing challenges.


\begin{result}
  \textbf{RQ2:} Different mutation operators result in different mutant
survivability; for example, the survival rate of LCR is almost twice as high as that of UOI.
\end{result}

\begin{figure}[tb]
\centering
\subfloat[Productivity per programming language.]{
\includegraphics[trim=0 20 0 0,clip,width=.4\textwidth]{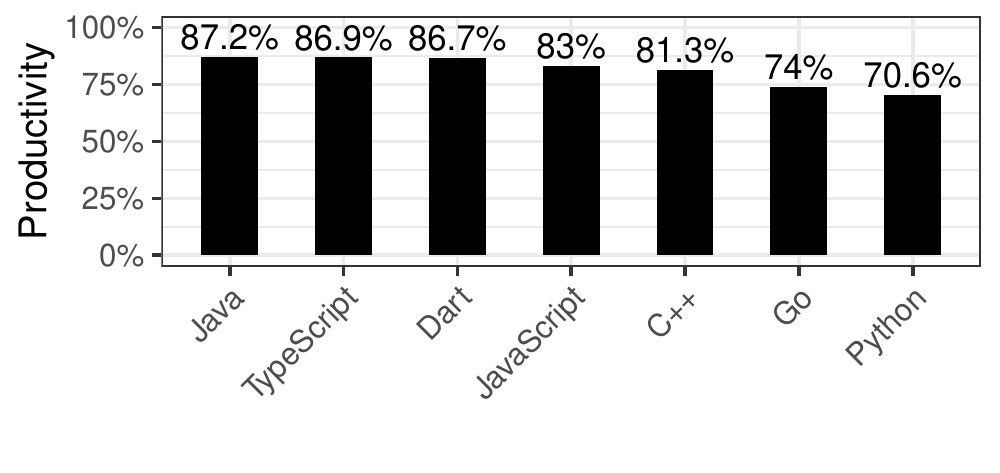}
}

\subfloat[Productivity per mutation operator.]{
\includegraphics[trim=0 20 0 0,clip,width=.4\textwidth]{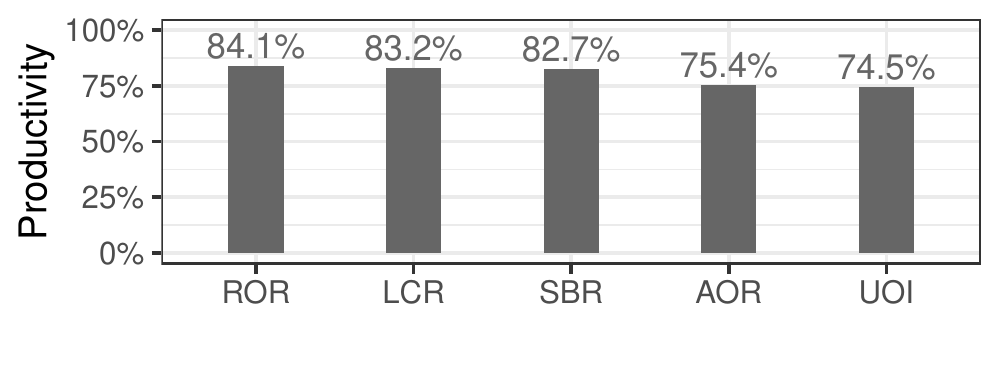}
}
\caption{Mutant productivity.}
\label{fig:usefulness}
\end{figure}

\subsection{RQ3 Mutant Productivity}
\label{results_feedback}

Mutant productivity is the most important measure, because it directly measures
the utility of a surfaced mutant.
Since we only generate a single mutant in a line, that mutant ideally should not just
survive the test suite but also be productive, allowing developers to improve the test suite
or the source code itself.
Given Google's high accuracy and actionability
requirements for surfacing code findings during code reviews, we rely on
developer feedback as the best available measure for mutant
productivity. Specifically, we consider a mutant a developer marked with
\textit{Please fix} to be more productive than others. Likewise, we consider a mutant
a developer marked with \textit{Not useful} to be less productive than others. (Note
that we excluded mutants for which no developer feedback is available from the analysis.)
We compare the mutant productivity across mutation operators and programming
languages.

Figure~\ref{fig:usefulness} shows the results, indicating that mutant
productivity is similar across mutation operators, with AOR and UOI mutants
being noticeably less productive. For example, ROR mutants
are productive 84.1\% of the time, whereas, UOI mutants are only productive 74.5\% of the time.
The differences between programming languages are even more pronounced, with
Java mutants being productive 87.2\% of the time, compared to Python mutants
that are productive 70.6\% of the time. This could be due to code conventions,
language common usecase scenarios, testing frameworks or simply the lack of heuristics.
We have found that Python code generally requires more tests because of the lack of the compiler.

\begin{result}
  \textbf{RQ3:} ROR, LCR, and SBR mutants show similar productivity, whereas AOR
  and  UOI mutants show noticeably lower productivity.
\end{result}

\subsection{RQ4 Mutation Context}
\label{results_context}

We examine whether context-based selection of mutation operators improves mutant
survivability and productivity. Specifically, we determine whether context-based
selection of mutation operators increases the probability of a generated mutant to survive and to
result in a \PleaseFix request, when compared to the random-selection baseline.

\begin{figure}[tb]
\centering
\includegraphics[trim=0 20 0 0,clip,width=\linewidth]{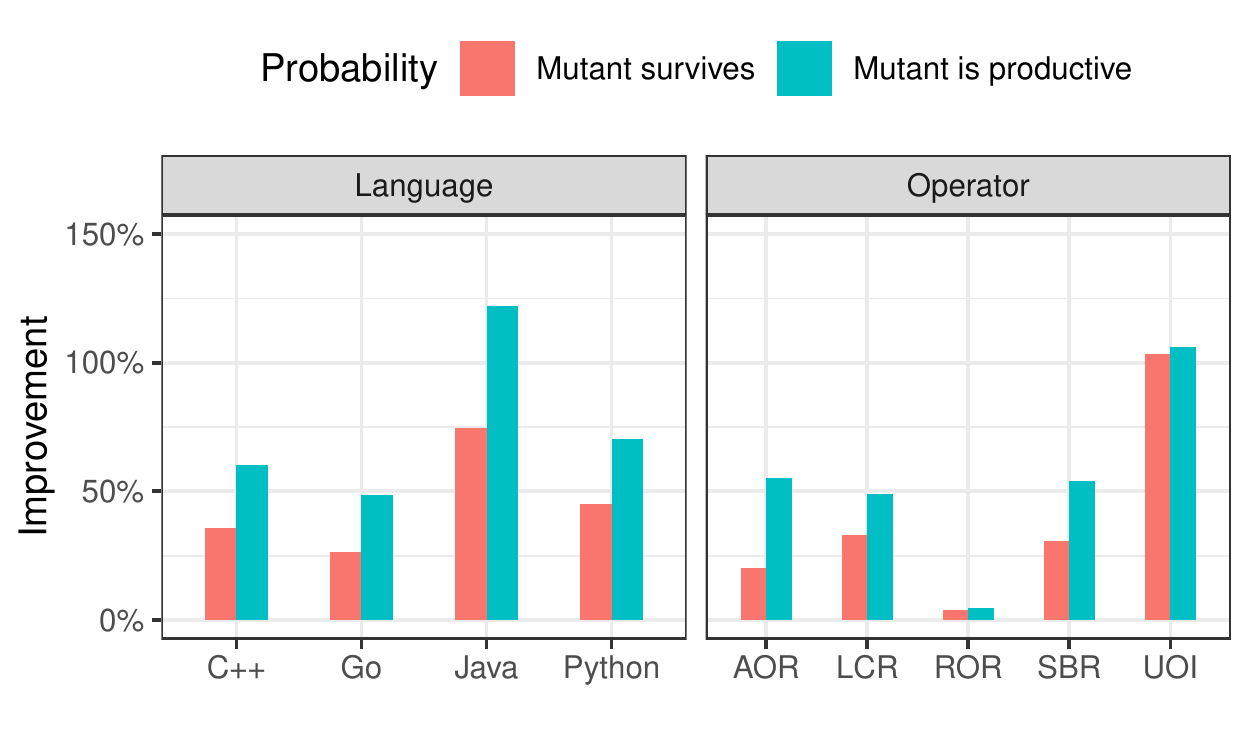}

\caption{Improvements achieved by
context-based selection.
(0\% improvement corresponds to random selection.)}
\label{fig:improvements}
\end{figure}

Figure~\ref{fig:improvements} shows that selecting mutation operators based on
the AST context of the node under mutation substantially increases the
probability of the generated mutant to survive and to result in a \PleaseFix
request. While improvements vary across programming languages and across mutation
operators, the context-based selection consistently outperforms random selection.
The largest productivity improvements are achieved for UOI, AOR, and SBR,
which generate most of all mutants. Intuitively, these improvements mean that
context-based selection results in twice as many productive UOI mutants (out of
all generated mutants), when
compared to random selection. Figure~\ref{fig:improvements} also shows to what
extent these improvements can be attributed to the fact that simply more mutants
are surfaced. Since the improvements for productivity increase even more than those
for survivability, context-based selection not only results in more surfaced
mutants but also in higher productivity of the surviving mutants.
Overall, the survival rate increases by over 40\% 
and the probability that a reviewer asks for a generated mutant to be fixed
increases by almost 50\%.

It is important to put these improvements into context. Probabilistic diff-based
mutation analysis aggressively trims down the number of considered mutants from
thousands in a representative file to a mere few, and enables mutants to be
effectively presented to developers as potential test targets.
The random-selection approach produces fewer surviving mutants of lower
productivity.

\begin{result}
	\textbf{RQ4:} Context-based selection improves the probability that a
generated mutant survives by more than 40\% and the probability that a generated mutant is
productive by almost 50\%.
\end{result}

\section{Related Work}
\label{sec:related}

There are several veins of research that are related to this work.
Just \etal proposed an AST-based program context model for predicting
mutant effectiveness~\cite{JustKA2017}.
Fernandez \etal developed various rules for Java programs to detect equivalent
and redundant mutants~\cite{FernandesUselessMutantsGPCE2017}.
The initial results are promising for developing selection strategies that
outperform random selection.
Further, Zhang~\etal used machine learning to predict mutation scores, both on
successive versions of a given project, and across
projects~\cite{ZhangPredictiveMutation2016}.
Finally, the PIT project makes mutation testing usable by practicing developers
and has gained adoption in the industry~\cite{PITWeb}.

There has been a lot of focus on computational costs and the equivalent
mutant problem~\cite{JiaH2011}.
There is much focus on avoiding redundant mutants, which leads to increase of
computational costs and inflation of the mutation score~\cite{JustS2015},
and instead favoring hard-to-detect mutants~\cite{YaoStubbornICSE14,Visser2016} or
dominator mutants~\cite{MinimalMutantsICST2014}. Mutant subsumption graphs have
similar goals but mutant productivity is much more fuzzy than dominance or
subsumption.

Effectiveness for mutants is primarily defined in terms of redundacy and
equivalence. This approach fails to consider the notion that non-reduntant
mutants might be unproductive or that equivalent mutants can be
productive~\cite{IneffectiveMutants2017}. From our experience, reporting 
equivalent mutants has been a vastly easier problem than reporting unproductive
non-reduntant and non-equivalent mutants.

Our approach for targeted mutant selection (Section~\ref{sec:criteria})
compares the context of mutants using tree hashes. The specific implementation
was driven by the need for consistency and efficiency, in order to make it
possible to look up similar AST contexts in real time during mutant creation.
In particular, the hash distances need to be preserved over time to improve the
targeted selection.
There are approaches to software clone detection~\cite{roy2007survey} that similarly use tree-distances (e.g.,~\cite{baxter1998clone,yang1991identifying,jiang2007deckard,wahler2004clone,evans2009clone}). Whether alternative distance measurements can be scaled for application at Google and whether they can further improve the targeted selection remains to be determined in future work.

This approach is similar to tree-based approaches
(e.g.,~\cite{baxter1998clone,yang1991identifying,jiang2007deckard,wahler2004clone,evans2009clone}) in software clone detection~\cite{roy2007survey}, which aims
to detect that a code fragment is a copy of some original code, with or without
modification. The AST-based techniques can detect additional categories of
modifications like identifier name changes or type aliases, that token-based
detection cannot, and the insensitivity of to variable names is important for
the mutation context. However, clone detection differs drastically in its goal:
it cares about detecting code with the same semantics, in spite of the syntactical
changes made to it. While clone detection might want to detect that an algorithm
has been copied and then changed slightly, e.g., a recursion rewritten to an equivalent
iterative algorithm, mutation testing context cares only about the neighboring AST nodes:
in the iterative algorithm, the most productive mutants will be those that thrived
before in such code, not the ones that thrived for a recursive algorithm.
In order to look up similar AST contexts in real time, as mutants are created, we
require a fast method that preserves hash distance over time. For these consistency
and efficiency reasons, we opted for the described tree-hashing approach.

\section{Conclusions}\label{conclusion}

Mutation testing has the potential to effectively guide
software testing and advance software quality. However, many mutants represent
unproductive test goals; writing tests for them does not improve test
suite efficacy and, even worse, negatively affects test maintainability.

Over the past six years, we have developed a scalable mutation testing approach
and mutant suppression rules that increased the ratio of productive mutants, as judged by developers, from
\prodStartPhaseOne to \prodEndPhaseTwo at Google.
Three strategies were key to success.
First, we devised an incremental mutation testing strategy, reporting at most one mutant per line of code---targeting lines that are changed and covered.
Second, we have created a set of rule-based heuristics for mutant suppression, based on developer feedback and manual analyses.
Third, we devised a probabilistic, targeted mutant selection approach that considers mutation context and historical mutation results.


Given the success of our mutation testing approach and the positive developer feedback, we are
planning to deploy it company-wide. We expect that this step will result in
additional refinements of the suppression and selection strategies in order to
maintain a mutant productivity rate around 90\%.
%
Furthermore, we will investigate the long-term effects of mutation testing on
developer behavior when writing tests as part of our future work.




\ifCLASSOPTIONcaptionsoff
  \newpage
\fi

\balance



\bibliographystyle{IEEEtran}
\bibliography{IEEEabrv,bib/bib-abbrv,bib/hash,bib/google,bib/mutation,bib/code2vec,bib/rjust}

\begin{thebibliography}{10}
\providecommand{\url}[1]{#1}
\csname url@samestyle\endcsname
\providecommand{\newblock}{\relax}
\providecommand{\bibinfo}[2]{#2}
\providecommand{\BIBentrySTDinterwordspacing}{\spaceskip=0pt\relax}
\providecommand{\BIBentryALTinterwordstretchfactor}{4}
\providecommand{\BIBentryALTinterwordspacing}{\spaceskip=\fontdimen2\font plus
\BIBentryALTinterwordstretchfactor\fontdimen3\font minus
  \fontdimen4\font\relax}
\providecommand{\BIBforeignlanguage}[2]{{%
\expandafter\ifx\csname l@#1\endcsname\relax
\typeout{** WARNING: IEEEtran.bst: No hyphenation pattern has been}%
\typeout{** loaded for the language `#1'. Using the pattern for}%
\typeout{** the default language instead.}%
\else
\language=\csname l@#1\endcsname
\fi
#2}}
\providecommand{\BIBdecl}{\relax}
\BIBdecl

\bibitem{IvankovicPJF2019}
M.~Ivankovi{\'c}, G.~Petrovi{\'c}, R.~Just, and G.~Fraser, ``Code coverage at
  google,'' in \emph{Proc. of ESEC/FSE}, August~26--30 2019, pp. 955--963.

\bibitem{offutt1996subsumption}
A.~J. Offutt and J.~M. Voas, ``Subsumption of condition coverage techniques by
  mutation testing,'' \emph{Department of Information and Software Systems
  Engineering, George Mason University, Tech. Rep. ISSE-TR-96-100}, 1996.

\bibitem{checkedcoverage}
D.~{Schuler} and A.~{Zeller}, ``Assessing oracle quality with checked
  coverage,'' in \emph{2011 Fourth IEEE International Conference on Software
  Testing, Verification and Validation}, 2011, pp. 90--99.

\bibitem{demillo1978hints}
R.~A. DeMillo, R.~J. Lipton, and F.~G. Sayward, ``Hints on test data selection:
  Help for the practicing programmer,'' \emph{Computer}, vol.~11, no.~4, pp.
  34--41, 1978.

\bibitem{andrews2006using}
J.~H. Andrews, L.~C. Briand, Y.~Labiche, and A.~S. Namin, ``Using mutation
  analysis for assessing and comparing testing coverage criteria,'' \emph{IEEE
  Transactions on Software Engineering}, vol.~32, no.~8, pp. 608--624, 2006.

\bibitem{just2014mutants}
R.~Just, D.~Jalali, L.~Inozemtseva, M.~D. Ernst, R.~Holmes, and G.~Fraser,
  ``Are mutants a valid substitute for real faults in software testing?'' in
  \emph{Proceedings of the International Symposium on Foundations of Software
  Engineering}.\hskip 1em plus 0.5em minus 0.4em\relax ACM, 2014, pp. 654--665.

\bibitem{ChenGTEHFAJ2020}
Y.~T. Chen, R.~Gopinath, A.~Tadakamalla, M.~D. Ernst, R.~Holmes, G.~Fraser,
  P.~Ammann, and R.~Just, ``Revisiting the relationship between fault
  detection, test adequacy criteria, and test set size,'' in \emph{Proc. of
  ASE}, September~21--25 2020, pp. 237--249.

\bibitem{45424}
R.~Potvin and J.~Levenberg, ``Why {Google} stores billions of lines of code in
  a single repository,'' \emph{Communications of the ACM}, vol.~59, pp. 78--87,
  2016.

\bibitem{swampUP}
``{How DevOps Accelerates "Ideas to Prod" at Google},''
  \url{https://swampup.jfrog.com/}.

\bibitem{SchulerZellerUncoveringICST2010}
D.~Schuler and A.~Zeller, ``(un-)covering equivalent mutants,'' in \emph{Proc.
  of ICST}, April 2010, pp. 45--54.

\bibitem{PetrovicIKAJ2018}
G.~Petrovi{\'c}, M.~Ivankovi{\'c}, B.~Kurtz, P.~Ammann, and R.~Just, ``An
  industrial application of mutation testing: Lessons, challenges, and research
  directions,'' in \emph{Proc. of Mutation}, Apr. 2018, pp. 47--53.

\bibitem{state_of_mt_at_google}
G.~Petrovic and M.~Ivankovic, ``State of {Mutation} {Testing} at {Google},'' in
  \emph{Proceedings of the 40th International Conference on Software
  Engineering 2017 (SEIP)}, 2018.

\bibitem{offutt2001mutation}
A.~J. Offutt and R.~H. Untch, ``Mutation 2000: Uniting the orthogonal,''
  \emph{Mutation testing for the new century}, pp. 34--44, 2001.

\bibitem{bazel}
``{Bazel build system},'' \url{https://bazel.io/}, 2015.

\bibitem{offutt1996experimental}
A.~J. Offutt, A.~Lee, G.~Rothermel, R.~H. Untch, and C.~Zapf, ``An experimental
  determination of sufficient mutant operators,'' \emph{ACM Transactions on
  Software Engineering and Methodology (TOSEM)}, vol.~5, no.~2, pp. 99--118,
  1996.

\bibitem{43322}
C.~Sadowski, J.~van Gogh, C.~Jaspan, E.~Soederberg, and C.~Winter, ``Tricorder:
  Building a program analysis ecosystem,'' in \emph{Software Conference (ICSE),
  2015}, 2015.

\bibitem{muchnick1997advanced}
S.~S. Muchnick, \emph{Advanced compiler design implementation}.\hskip 1em plus
  0.5em minus 0.4em\relax Morgan Kaufmann, 1997.

\bibitem{grpc}
G.~Inc., ``{gRPC: A high performance, open-source universal RPC framework},''
  \url{https://grpc.io}, 2006.

\bibitem{tatikonda2010hashing}
S.~Tatikonda and S.~Parthasarathy, ``Hashing tree-structured data: Methods and
  applications,'' in \emph{2010 IEEE 26th International Conference on Data
  Engineering (ICDE 2010)}.\hskip 1em plus 0.5em minus 0.4em\relax IEEE, 2010,
  pp. 429--440.

\bibitem{minhash}
A.~Z. Broder, M.~Charikar, A.~M. Frieze, and M.~Mitzenmacher, ``Min-wise
  independent permutations,'' \emph{Journal of Computer and System Sciences},
  vol.~60, no.~3, pp. 630--659, 2000.

\bibitem{googlestyleguide}
``{Google Style Guides},'' \url{https://google.github.io/styleguide/}.

\bibitem{bacchelli2013expectations}
A.~Bacchelli and C.~Bird, ``Expectations, outcomes, and challenges of modern
  code review,'' in \emph{2013 35th International Conference on Software
  Engineering (ICSE)}.\hskip 1em plus 0.5em minus 0.4em\relax IEEE, 2013, pp.
  712--721.

\bibitem{DeMillo:1978:HTD:1300736.1301357}
R.~A. DeMillo, R.~J. Lipton, and F.~G. Sayward, ``Hints on test data selection:
  Help for the practicing programmer,'' \emph{Computer}, vol.~11, no.~4, pp.
  34--41, Apr. 1978.

\bibitem{JustKA2017}
R.~Just, R.~J. Kurtz, and P.~Ammann, ``Inferring mutant utility from program
  context,'' in \emph{Proc. of ISSTA}, July 2017, pp. 284--294.

\bibitem{FernandesUselessMutantsGPCE2017}
L.~Fernandes, M.~Ribeiro, L.~Carvalho, R.~Gheyi, M.~Mongiovi, A.~Santos,
  A.~Cavalcanti, F.~Ferrari, and J.~C. Maldonado, ``Avoiding useless mutants,''
  in \emph{Proc. of GPCE}, October 2017, pp. 187--198.

\bibitem{ZhangPredictiveMutation2016}
J.~Zhang, Z.~Wang, L.~Zhang, D.~Hao, L.~Zang, S.~Cheng, and L.~Zhang,
  ``Predictive mutation testing,'' in \emph{Proc. of ISSTA}, July 2016, pp.
  342--353.

\bibitem{PITWeb}
H.~Coles, ``Real world mutation testing,'' http://pitest.org, last accessed
  January 2018.

\bibitem{JiaH2011}
Y.~Jia and M.~Harman, ``An analysis and survey of the development of mutation
  testing,'' \emph{IEEE TSE}, vol.~37, no.~5, pp. 649--678, 2011.

\bibitem{JustS2015}
R.~Just and F.~Schweiggert, ``Higher accuracy and lower run time: efficient
  mutation analysis using non-redundant mutation operators,'' \emph{JSTVR},
  vol.~25, no. 5-7, pp. 490--507, 2015.

\bibitem{YaoStubbornICSE14}
X.~Yao, M.~Harman, and Y.~Jia, ``A study of equivalent and stubborn mutation
  operators using human analysis of equivalence,'' in \emph{Proc. of ICSE}, May
  2014, pp. 919--930.

\bibitem{Visser2016}
W.~Visser, ``What makes killing a mutant hard,'' in \emph{Proc. of ASE},
  September 2016, pp. 39--44.

\bibitem{MinimalMutantsICST2014}
P.~Ammann, M.~E. Delamaro, and J.~Offutt, ``Establishing theoretical minimal
  sets of mutants,'' in \emph{Proc. of ICST}, 2014, pp. 21--31.

\bibitem{IneffectiveMutants2017}
P.~McMinn, C.~J. Wright, C.~J. McCurdy, and G.~Kapfhammer, ``Automatic
  detection and removal of ineffective mutants for the mutation analysis of
  relational database schemas,'' \emph{IEEE TSE}, 2017.

\bibitem{roy2007survey}
C.~K. Roy and J.~R. Cordy, ``A survey on software clone detection research,''
  \emph{Queen’s School of Computing TR}, vol. 541, no. 115, pp. 64--68, 2007.

\bibitem{baxter1998clone}
I.~D. Baxter, A.~Yahin, L.~Moura, M.~Sant'Anna, and L.~Bier, ``Clone detection
  using abstract syntax trees,'' in \emph{Proceedings. International Conference
  on Software Maintenance (Cat. No. 98CB36272)}.\hskip 1em plus 0.5em minus
  0.4em\relax IEEE, 1998, pp. 368--377.

\bibitem{yang1991identifying}
W.~Yang, ``Identifying syntactic differences between two programs,''
  \emph{Software: Practice and Experience}, vol.~21, no.~7, pp. 739--755, 1991.

\bibitem{jiang2007deckard}
L.~Jiang, G.~Misherghi, Z.~Su, and S.~Glondu, ``Deckard: Scalable and accurate
  tree-based detection of code clones,'' in \emph{International Conference on
  Software Engineering (ICSE'07)}.\hskip 1em plus 0.5em minus 0.4em\relax IEEE,
  2007, pp. 96--105.

\bibitem{wahler2004clone}
V.~Wahler, D.~Seipel, J.~Wolff, and G.~Fischer, ``Clone detection in source
  code by frequent itemset techniques,'' in \emph{Source Code Analysis and
  Manipulation, Fourth IEEE International Workshop on}.\hskip 1em plus 0.5em
  minus 0.4em\relax IEEE, 2004, pp. 128--135.

\bibitem{evans2009clone}
W.~S. Evans, C.~W. Fraser, and F.~Ma, ``Clone detection via structural
  abstraction,'' \emph{Software Quality Journal}, vol.~17, no.~4, pp. 309--330,
  2009.

\end{thebibliography}
%
%
%

%

\begin{IEEEbiographynophoto}{Goran~Petrovi\'{c}}
Goran~Petrovi\'{c} is a Staff Software Engineer at Google Switzerland, Z{\"u}rich. He received an MS in Computer Science from University of Zagreb, Croatia, in 2009. His main research interests are software quality metrics and improvements, ranging from prevention of software defects to evaluation of software design reusability and maintenance costs and automated large scale software refactoring.
\end{IEEEbiographynophoto}

\begin{IEEEbiographynophoto}{Marko~Ivankovi\'{c}}
  Marko Ivankovi\'{c} is a Staff Software Engineer at Google Switzerland, Z{\"u}rich. He received an MS in Computer Science from University of Zagreb, in 2011. His work focuses on Software Engineering as a discipline, large scale code base manipulation, code metrics and developer workflows.
\end{IEEEbiographynophoto}

\begin{IEEEbiographynophoto}{Gordon Fraser}
Gordon Fraser is a full professor in Computer Science at the University of Passau, Germany. He received a PhD in computer science from Graz University of Technology, Austria, in 2007, worked as a post-doc at Saarland University, and was a Senior Lecturer at the University of Sheffield, UK. The central theme of his research is improving software quality, and his recent research concerns the prevention, detection, and removal of defects in software. 
\end{IEEEbiographynophoto}

\begin{IEEEbiographynophoto}{Ren{\'e} Just}
Ren{\'e} Just is an Assistant Professor at the University of Washington. His
research interests are in software engineering, software security, and data
science, in particular static and dynamic program analysis, mobile security, and
applied statistics and machine learning. He is the recipient of an NSF CAREER
Award, and his research in the area of software engineering won three ACM
SIGSOFT Distinguished Paper Awards. He develops research and educational
infrastructures that are widely adopted by other researchers and instructors
(e.g., Defects4J and the Major mutation framework).
\end{IEEEbiographynophoto}





\ifTR
\newpage
\appendices
\lstset{
    autogobble,
    columns=fullflexible,
    showspaces=false,
    showtabs=false,
    breaklines=true,
    showstringspaces=false,
    breakatwhitespace=true,
    escapeinside={(*|}{|*)},
    commentstyle=\color{greencomments},
    keywordstyle=\color{bluekeywords},
    stringstyle=\color{redstrings},
    numberstyle=\color{graynumbers},
    basicstyle=\ttfamily\footnotesize,
    frame=l,
    framesep=12pt,
    xleftmargin=12pt,
    tabsize=4,
    captionpos=b
}

\lstdefinestyle{diffs}{
  moredelim=**[is][\ttfamily\footnotesize\color{red}]{|}{|},
  moredelim=**[is][\ttfamily\footnotesize\color{green}]{H}{H},
}

\section{Arid Node Heuristics}
\label{appendix:heuristics}

Nodes of the abstract syntax tree (AST) are \emph{arid} if applying mutation operators on them or their subtrees would lead to unproductive mutants. An unproductive mutant is either trivially equivalent to the original program, or if it is detectable then adding a test for it would not lead to an actual improvement of the test suite. The decision of whether a node of the AST is arid is implemented using heuristics built on developer feedback over time. In general, these heuristics are specifically tailored for the environment of the developers who provided the feedback, and a different context will require deriving new, appropriate heuristics. In this appendix, we summarize the main categories of such heuristics. We first summarize the main categories of arid node heuristics that are indendent of a specific programming language, then we describe heuristics developed specifically for different programming languages. For each heuristic, we provide examples of unproductive mutants that the heuristic addresses.

\subsection{Language Independent Heuristics}

\subsubsection{Logging Frameworks}

Logging statements are rarely tested outside of the code of the logging systems
themselves. Mutants in logging statements are usually unproductive and would not lead to tests that improve software quality.

\begin{figure}[H]
\begin{lstlisting}[language=go,style=diffs]
LOG(INFO) << "Duration: " << (absl::Now() |-| start);
\end{lstlisting}
\vfill\hrule\vfill
\begin{lstlisting}[language=go,style=diffs]
LOG(INFO) << "Duration: " << (absl::Now() H+H start);
\end{lstlisting}
\end{figure}

A special case of the logging statement heuristic concerns the \texttt{Console}
class available in the browser that can be used for logging; mutants in that
code are unproductive test goals.

\begin{figure}[H]
\begin{lstlisting}[language=Java,style=diffs]
console.log('duration is ', new Date() |-| start);
\end{lstlisting}
\vfill\hrule\vfill
\begin{lstlisting}[language=Java,style=diffs]
console.log('duration is ', new Date() H+H start);
\end{lstlisting}
\end{figure}

Similar is true for other console methods like assert.

\textbf{Implementation.} This is implemented using AST-level arid node tagging, matching call expression or
macros.

\textbf{Soundness.} This heuristic is sound when applied to source code that
does not explicitly test the logging code itself, which is easy to detect using
the build system.

\subsubsection{Memory and Capacity Functionality}
Often it makes sense to pre-allocate memory for efficiency, when the total size
is known in advance. Mutants in these memory size specifications are not good
test goals; they usually create functionally equivalent code and are not
killable by standard testing methods.

\begin{figure}[H]
\begin{lstlisting}[language=go,style=diffs]
std::vector<std::string> merged(left.size() H+H right.size());
absl::c_copy(left, std::back_inserter(merged));
absl::c_copy(right, std::back_inserter(merged));
\end{lstlisting}
\vfill\hrule\vfill
\begin{lstlisting}[language=go,style=diffs]
std::vector<std::string> merged(left.size() |-| right.size());
absl::c_copy(left, std::back_inserter(merged));
absl::c_copy(right, std::back_inserter(merged));
\end{lstlisting}
\end{figure}

In this example, the only consequence will be that the vector may need to grow itself and that
will take extra time. The same also holds for Java collections, e.g.,

\begin{figure}[H]
\begin{lstlisting}[language=Java,style=diffs]
List<String> merged = new ArrayList<>(left.length() |+| right.length());
\end{lstlisting}
\vfill\hrule\vfill
\begin{lstlisting}[language=Java,style=diffs]
List<String> merged = new ArrayList<>(left.length() H-H right.length());
\end{lstlisting}
\end{figure}

\begin{figure}[H]
\begin{lstlisting}[language=Java,style=diffs]
List<String> merged = Lists.newArrayListWithCapacity(left.length() |+| right.length());
\end{lstlisting}
\vfill\hrule\vfill
\begin{lstlisting}[language=Java,style=diffs]
List<String> merged = Lists.newArrayListWithCapacity(left.length() H-H right.length());
\end{lstlisting}
\end{figure}

Similar constructs exist in all programming languages, and the heuristic extends to all of these such as \texttt{std::vector::resize}, or
\texttt{reserve}, \texttt{shrink\_to\_fit}, \texttt{free}, \texttt{delete}.
These represent a family of common functions of many containers in many
languages, \texttt{std::vector} being just a representative example.

Another interesting example are cache prefetch instructions added with SSE,
\texttt{prefetch0}, \texttt{prefetch}, \texttt{prefetch2} and
\texttt{prefetchnta} accessible with \texttt{\_\_builtin} or
directly by an asm block.

\begin{figure}[H]
\begin{lstlisting}[language=C,style=diffs]
-  |__builtin_prefetch(&obj, 0, 3);|
\end{lstlisting}
\end{figure}

\textbf{Implementation.} This is implemented using AST-level arid node tagging, matching call expressions.

\textbf{Soundness.} This heuristic is sound; it uses exact symbols and type
names.

\subsubsection{Monitoring Systems} Although it may be debatable whether monitoring logic should be tested or not, developers did not use such mutants productively and instead reported them as being unproductive. Consequently, heuristics mark AST nodes related to monitoring logic as arid.

\begin{figure}[H]
\begin{lstlisting}[language=go,style=diffs]
#include <prometheus/counter.h>

auto& counter_family = prometheus::BuildCounter().Name("time").Register(*r);
auto& error_counter = counter_family.Add({{"error", "value"}}});

error_counter.Increment(run1.errors().size() |+| run2.errors().size());
\end{lstlisting}
\vfill\hrule\vfill
\begin{lstlisting}[language=go,style=diffs]
#include <prometheus/counter.h>

auto& counter_family = prometheus::BuildCounter().Name("time").Register(*r);
auto& error_counter = counter_family.Add({{"error", "value"}}});

error_counter.Increment(run1.errors().size() H-H run2.errors().size());
\end{lstlisting}
\end{figure}

\textbf{Implementation.} This is implemented using AST-level arid node tagging, matching constructor or call
expressions.

\textbf{Soundness.} This heuristic is sound; it uses exact symbols and type
names.

\subsubsection{Time Related Code} Clocks are usually faked in tests, and networking
calls are short-circuited to special RPC implementations for testing; it therefore rarely makes sense to mutate time expressions when used in a deadline-context, because they would lead to unproductive mutants.

\begin{figure}[H]
\begin{lstlisting}[language=go,style=diffs]
-  |::SleepFor(absl::Seconds(5));|
\end{lstlisting}
\end{figure}

The same holds for other types of network-code, such as setting deadlines:

\begin{figure}[H]
\begin{lstlisting}[language=go,style=diffs]
context.set_deadline(std::chrono::system_clock::now() |+| std::chrono::milliseconds(10));
\end{lstlisting}
\vfill\hrule\vfill
\begin{lstlisting}[language=go,style=diffs]
context.set_deadline(std::chrono::system_clock::now() H-H std::chrono::milliseconds(10));
\end{lstlisting}
\end{figure}

\textbf{Implementation.} This is implemented using AST-level arid node tagging, matching constructor or call
expressions.

\textbf{Soundness.} This heuristic is sound; it uses exact symbols and type
names.

\subsubsection{Tracing and Debugging} Code is often adorned with debugging and
tracing information that may be even excluded in the release builds, but present
while testing. This code serves its purpose, but is usually impossible to test
and mutants in that code do not make good test goals.

\begin{figure}[H]
\begin{lstlisting}[language=c,style=diffs]
-  |ASSERT_GT(input.size(), 0);|
\end{lstlisting}
\end{figure}

\begin{figure}[H]
\begin{lstlisting}[language=c,style=diffs]
-  |assert(x != nullptr);|
\end{lstlisting}
\end{figure}

\begin{figure}[H]
\begin{lstlisting}[language=c,style=diffs]
-  |TRACE(x);|
\end{lstlisting}
\end{figure}

\begin{figure}[H]
\begin{lstlisting}[language=Java,style=diffs]
-  |Preconditions.checkNotNull(v);|
\end{lstlisting}
\end{figure}

\begin{figure}[H]
\begin{lstlisting}[language=Java,style=diffs]
-  |exception.printStackTrace();|
\end{lstlisting}
\end{figure}

In general, check-failures usually make the program segfault and serve as a
last line of defense, and tracing is used for debugging purposes, and so neither results in productive mutants.

\textbf{Implementation.} This is implemented using AST-level arid node tagging, matching constructors, call
expressions or macros.

\textbf{Soundness.} This heuristic is sound; it uses exact symbols and type
names.

\subsubsection{Programming Model Frameworks} There are specialized frameworks for
specifying complex work conceptually and then executing that work in a different
way, where the code that is written serves as a model for the intent, not the
real logic that gets executed. Some examples of this principle are Apache Beam
and TensorFlow.

\begin{figure}[H]
\begin{lstlisting}[language=Java,style=diffs]
Pipeline p = Pipeline.create(options);
PCollection<String, Long> word_counts = p
  .apply(TextIO.read().from(options.getInputFile()))
  .apply("ExtractWords", new WordExtractor())
  .apply(Count.<String>perElement())
  .apply("FormatResults", new ResultFormatter());

// materializes the results
-  |PipelineRunner.run(p);|
\end{lstlisting}
\end{figure}

In this example, developers usually test the components of the pipeline, but not the code assembling the pipeline. Similar examples exist for TensorFlow:

\begin{figure}[H]
\begin{lstlisting}[language=Python,style=diffs]
- |tf.compat.v1.enable_eager_execution()|
  assert tf.multiply(6, 7).numpy() == 42
\end{lstlisting}
\end{figure}

\textbf{Implementation.} This is implemented using AST-level arid node tagging, matching constructor or call
expressions.

\textbf{Soundness.} This heuristic is not sound. Because it is based on
best-effort matching of code structures that look arid and often are, it can
suppress productive mutants.

\subsubsection{Block Body Uncovered}

Suppose that a block entry condition (e.g., of an \texttt{if}-statement) is covered by tests, but the condition is not fulfilled by any tests and thus the corresponding block is not covered. Most mutants of the condition would only help the developers to identify that no test covers the relevant branch yet. However, the same information is already provided by coverage, and so mutants in such \texttt{if}-conditions are deemed unproductive.
Mutants like this can indeed inform about test suite quality, but coverage is a far simpler test goal for the developers to act on in this case, and for that reason we use coverage to drive test case implementation, and mutatns for their subsequent improvement.

\textbf{Implementation.} This is implemented using AST-level arid node tagging, aided by line code
coverage data.

\textbf{Soundness.} While most mutants are indeed unproductive, the heuristic is not entirely sound as there may be mutants that reveal information about boundary cases of the condition.
Since coverage points out that a branch is not taken, forcing boundary-check tests prior to even covering both branches is pre-mature; if the tests written for the coverage test goal
do not check boundary conditions, mutants can then be reported as new test goals.

\subsubsection{Arithmetic Operator with a no-op Child}
In some cases, mostly due to style, code will be written with explicit zeros for
some parts of an expression. For example:

\begin{figure}[H]
\begin{lstlisting}[language=go,style=diffs]
data[i] |+ 0 * sizeof(char)|, data[i] + 4 * sizeof(char), data[i] + 8 * sizeof(char);
\end{lstlisting}
\end{figure}

Mutating the binary operator \texttt{+} by removing the right-hand side (the \texttt{0 *
sizeof(char)}), leaving only left-hand side of the binary operator
 (the \texttt{data[i]}), results in an equivalent mutant. The code is
simply written in such a way because it deals with low-level instructions and the code style requires that each offset be explicitly written, and all lines
equally aligned so each offset is at the same column.

\textbf{Implementation.} This is implemented using AST-level arid node tagging, matching expressions.

\textbf{Soundness.} This heuristic is sound because it has the full type and
expression information available.

\subsubsection{Logical Comparator of POD with Zero Values} When comparing a
plain-old-data structure with its zero-value, there is a possibility for
creating an equivalent mutant. For example, a conditional statement \texttt{if
(x != 0)}, with \texttt{x} having a primitive or record type, is equivalent to
\texttt{if (x)}. In that case, mutating the condition \texttt{x != 0} to the
left-hand-side operand \texttt{x} produces an equivalent mutant.

\begin{figure}[H]
\begin{lstlisting}[language=go,style=diffs]
if (|x !== 0|) {
  return 5;
}
\end{lstlisting}
\vfill\hrule\vfill
\begin{lstlisting}[language=go,style=diffs]
if (HxH) {
  return 5;
}
\end{lstlisting}
\end{figure}

\textbf{Implementation.} This is implemented using AST-level arid node tagging, matching expressions.

\textbf{Soundness.} This heuristic is sound when full type and expression information is available.

\subsubsection{Logical Comparator with Null Child}
When comparing something to \texttt{nullptr} and its corresponding value in other languages (\texttt{NULL}, \texttt{nil}, \texttt{null}, \texttt{None}, ...),
picking the left (or right, depending where the null value is) is equivalent to
replacing the binary operator with false.

\begin{figure}[H]
\begin{lstlisting}[language=go,style=diffs]
if (|worker_ == nullptr|)
\end{lstlisting}
\vfill\hrule\vfill
\begin{lstlisting}[language=go,style=diffs]
+ if (HnullptrH)  // `if (false)` is the equivalent mutation
\end{lstlisting}
\end{figure}

In an expression of format \texttt{x != nullptr}, mutating it to \texttt{x} is
an equivalent mutant.

\begin{figure}[H]
\begin{lstlisting}[language=go,style=diffs]
if (|worker_ != nullptr|) worker_->DoWork();
\end{lstlisting}
\vfill\hrule\vfill
\begin{lstlisting}[language=go,style=diffs]
if (Hworker_H) worker_->DoWork();
\end{lstlisting}
\end{figure}

\textbf{Implementation.} This is implemented using AST-level arid node tagging, matching expressions.

\textbf{Soundness.} This heuristic is sound because it has the full type and expression information available.

\subsubsection{Floating Point Equality}
Floating point equality comparison, except for special values such as zero, is mostly meaningless. For a number \texttt{x} that is not 0, replacing \texttt{f() > x} with \texttt{f() >= x} is not a good test goal.
\begin{figure}[H]
\begin{lstlisting}[language=go,style=diffs]
return normalized_score |>| 0.95
\end{lstlisting}
\vfill\hrule\vfill
\begin{lstlisting}[language=go,style=diffs]
return normalized_score H>=H 0.95
\end{lstlisting}
\end{figure}

\textbf{Implementation.} This is implemented using AST-level arid node tagging, matching expressions.

\textbf{Soundness.} This heuristic is sound because it has the full type and
expression information available.

\subsubsection{Expression and Statement Deletion} Many statements can be deleted,
but usually more cannot, if the code is to compile. This is obvious in itself,
but it is worth reporting general types of nodes that are best not deleted. Some
of them are: conditional (ternary) operator: \texttt{b} in \texttt{a ? b : c},
parameters of call expressions: \texttt{a} in \texttt{f(a)}, non-assignment
binary operators, unary operators that are not a standalone statement but within
a compound, return, label, default and declaration statements, blocks containing
a return path within non-void functions, only statements in non-void functions
(function with 1 statement). Some of these rules change from language to
language, or are applicable only in some languages, but the ideas carry. In C++,
one may have a function without a return statement and when compiled with the
right set of compiler flags, it compiles, but the return value is undefined, and
in some other languages it would fail to compile and no amount of compiler flags
could change that. Blocks can be deleted, or replaced with an empty block \texttt{\{\}},
or in Python a block with \texttt{pass}.

\textbf{Implementation.} This is implemented using AST-level arid node tagging, matching nodes.

\textbf{Soundness.} This heuristic is not sound because it might suppress some
mutants that would be productive.

\subsubsection{Program Flags}
Program flags, passed in as arguments and parsed by some flag framework like
Abseil, are a way to configure the program. Often, tests will inject the fake
flag values, but often they will ignore them; they may be used for algorithm
tweaking (max threads in pool, max size of cache, deadline for network
operations). Other flags will inform the program about the location of
dependencies on the network, or resources on the file system; these are usually
faked in tests and injected directly into the code using the programming API
rather than flags, since the code is directly invoked, rather than forked into.

\begin{figure}[H]
\begin{lstlisting}[language=go,style=diffs]
-  |flags.DEFINE_string('name', 'Jane Random', 'Your name.')|
\end{lstlisting}
\end{figure}

\begin{figure}[H]
\begin{lstlisting}[style=diffs]
flags.DEFINE_integer('stack_size', 1000 |*| 1000, 'Size of the call stack.')
\end{lstlisting}
\vfill\hrule\vfill
\begin{lstlisting}[style=diffs]
flags.DEFINE_integer('stack_size', 1000 H/H 1000, 'Size of the call stack.')
\end{lstlisting}
\end{figure}

\begin{figure}[H]
\begin{lstlisting}[style=diffs]
flags.DEFINE_integer('rpc_deadline_seconds', 5 |*| 60, 'Network deadline.')
\end{lstlisting}
\vfill\hrule\vfill
\begin{lstlisting}[style=diffs]
flags.DEFINE_integer('rpc_deadline_seconds', 5 H+H 60, 'Network deadline.')
\end{lstlisting}
\end{figure}

\textbf{Implementation.} This is implemented using AST-level arid node tagging, matching expressions.

\textbf{Soundness.} This heuristic is sound because it has the full type and
expression information available.

\subsubsection{Low-level APIs}
If the code directly accesses the operating system using the standard libraries
(glibc) or Python's os or shutil libraries (e.g., to copy some files, create a
directory, or to print on the screen), then the program is probably some kind of a utility script and mutating these calls results in unproductive mutants: these calls are hard to mock (except in Python) and mostly unproductive test targets. There are exceptions, e.g., an API that wraps this communication and is used by various projects, but for the most part there are few of those and many more of simple utility scripts for doing basic filesystem operations. We can be sure that these are not critical programs because the standard libraries cannot use any of the standard storages, just local disk, and are rarely used in production.

\begin{figure}[H]
\begin{lstlisting}[language=go,style=diffs]
-  |shutil.rmtree(dir)|
\end{lstlisting}
\end{figure}

\begin{figure}[H]
\begin{lstlisting}[language=go,style=diffs]
-  |os.rename(from, to)|
\end{lstlisting}
\end{figure}

\textbf{Implementation.} This is implemented using AST-level arid node tagging, matching expressions.

\textbf{Soundness.} This heuristic is sound because it has the full type and
expression information available.

\subsubsection{Stream Operations}
Streams like \texttt{stdout}, \texttt{stderr}, or any other cached buffer, flush when the buffer
fills to some point, or on special events. Removing the flush operations on
various streams should change no behavior from the test point of view, and therefore mutants of such statements are not productive test goals. The same also holds for close operations on files or other buffers.

\begin{figure}[H]
\begin{lstlisting}[language=c,style=diffs]
-  |buffer.flush();|
\end{lstlisting}
\end{figure}

\begin{figure}[H]
\begin{lstlisting}[language=c,style=diffs]
-  |file.close();|
\end{lstlisting}
\end{figure}

\textbf{Implementation.} This is implemented using AST-level arid node tagging, matching call expressions.

\textbf{Soundness.} This heuristic is not sound because there are conceivable
code constructs in which buffer operations change the perceived behavior (e.g.,
in concurrent stream manipulation).

\subsubsection{Gate Configuration}
It is very common to use flags or some other mechanisms to facilitate easy
switching between different implementations, or control the state of rollout.
Consider the following:

\begin{figure}[H]
\begin{lstlisting}[language=python,style=diffs]
class Controller(object):
  USE_NEXT_GEN_BACKEND = |True|
\end{lstlisting}
\vfill\hrule\vfill
\begin{lstlisting}[language=python,style=diffs]
class Controller(object):
  USE_NEXT_GEN_BACKEND = HFalseH
\end{lstlisting}
\end{figure}

In this example there are two implementations, an old and a new one, but ideally both should work
correctly, and then it becomes impossible to distinguish by tests that there is a difference.

Similarly, a more gradual approach might have something like this:

\begin{figure}[H]
\begin{lstlisting}[language=java,style=diffs]
private static final Double nextGenTrafficRatio = |0.1|;
\end{lstlisting}
\vfill\hrule\vfill
\begin{lstlisting}[language=java,style=diffs]
private static final Double nextGenTrafficRatio = H0.1 + 1H;
\end{lstlisting}
\end{figure}

Some ratio of traffic can exercise a new implementation, for easier incremental
control.  Mutants in such global switches, usually determinable from code style,
do not make for good test goals.

\textbf{Implementation.} This is implemented using AST-level arid node tagging, matching nodes.

\textbf{Soundness.} This heuristic is not sound because it is guessing the
meaning of a class field based on its value and location, and it might be wrong.

\subsubsection{Cached lookups}
Often, values are cached/memoized to avoid redundant recalculation. Removing
the cache lookup slows down the program, but functionally
does not change anything, producing an equivalent, and thus unproductive, mutant.

\begin{figure}[H]
\begin{lstlisting}[style=diffs]
  def fib(n):
-   |if n in cache:|
-     |return cache[n]|
    if n == 1:
      value = 1
    elif n == 2:
      value = 1
    elif n > 2:
      value = fib(n - 1) + fib(n -2)
    cache[n] = value
    return value
\end{lstlisting}
\end{figure}

\textbf{Implementation.} This is implemented using AST-level arid node tagging, matching complex code
structures. The code structure that is considered a cached lookup must
fulfill the following: a) it must lookup an input parameter in a dissociative
container and return from it under that key if found, b) it must store the value
that it otherwise returns in the same container under the same key.

\textbf{Soundness.} This heuristic is not sound because it only checks for
probable code structures.

\subsubsection{Infinity}
There are various representations of infinity in mathematical libraries in
various languages. Incrementing or decrementing these produces an equivalent, and thus unproductive, mutant.

\begin{figure}[H]
\begin{lstlisting}[language=go,style=diffs]
x = a.replace([|numpy.inf|, -numpy.inf])
\end{lstlisting}
\vfill\hrule\vfill
\begin{lstlisting}[language=go,style=diffs]
x = a.replace([Hnumpy.inf + 1H, -numpy.inf])
\end{lstlisting}
\end{figure}

\textbf{Implementation.} This is implemented using AST-level arid node tagging, matching expressions.

\textbf{Soundness.} This heuristic is sound because it has the full type and
expression information available.

\subsubsection{Insensitive Arguments}
There are some functions that are insensitive to precise values or use them as
an indication only. These, if mutated, should be mutated to a degree that they
change not only the value, but also the indication of that value. For example, in Python the \texttt{zip} builtin makes an iterator that aggregates elements from each of the
iterables passed to it. The iterator stops when the shortest input iterable is
exhausted, meaning that changing the size of one of the parameters is not
guaranteed to affect the result.

\begin{figure}[H]
\begin{lstlisting}[language=go,style=diffs]
zip(a[i:|j|], b[j:k], c[k:m])
\end{lstlisting}
\vfill\hrule\vfill
\begin{lstlisting}[language=go,style=diffs]
zip(a[i:Hj + 1H], b[j:k], c[k:m])
\end{lstlisting}
\end{figure}

Incrementing and decrementing indices within \texttt{zip} parameters has may likely create equivalent (unproductive) mutants. Another example
is given by comparator functions in any context: It is very common for comparators to take in two values, and return -1, 0 or 1, if one element is less than the other, equal to it or greater than it, in whatever semantics the author defines. Commonly, any negative value implies the former, and any positive value implies the latter, while zero implies equality. As an example, consider Java Collections:

\begin{figure}[H]
\begin{lstlisting}[language=go,style=diffs]
list.sort((Person p1, Person p2) -> p1.getAge() - p2.getAge());
\end{lstlisting}
\vfill\hrule\vfill
\begin{lstlisting}[language=go,style=diffs]
list.sort((Person p1, Person p2) -> p1.getAge() - p2.getAge() H+ 1H);
\end{lstlisting}
\end{figure}

This mutant can only be helpful when the age difference is exactly -1 or 0, for
any other combination it is an equivalent mutant and thus an unproductive test target. 

Another Java example is the \texttt{String::split} method, for which one of the overloaded versions takes two parameters, the regex to define the split and the limit that controls the number of times the pattern is applied, affecting the length of the resulting array. According to the API specification, if te limit is non-positive then the pattern will be applied as many times as possible. This means that any negative number has the same semantics.

\begin{figure}[H]
\begin{lstlisting}[language=go,style=diffs]
String[] parts = key.split(",", |-1|);
\end{lstlisting}
\vfill\hrule\vfill
\begin{lstlisting}[language=go,style=diffs]
String[] parts = key.split(",", H-2H);
\end{lstlisting}
\end{figure}

Finally, another example is a loop spec with a step. When changing the range condition,
it has to be changed at least the full step for the change to have an effect.

\begin{figure}[H]
\begin{lstlisting}[language=go,style=diffs]
x = l[1:|10 + 2 * 7|:14]
\end{lstlisting}
\vfill\hrule\vfill
\begin{lstlisting}[language=go,style=diffs]
x = l[1:H10 + 2 * 7 + 1H:14]
\end{lstlisting}
\end{figure}

\begin{figure}[H]
\begin{lstlisting}[language=go,style=diffs]
for (int i = 1; i < |10 + 2 * 7|; i += 14) { std::cout << i << std::endl; }
\end{lstlisting}
\vfill\hrule\vfill
\begin{lstlisting}[language=go,style=diffs]
for (int i = 1; i < H10 + 2 * 7 + 1H; i += 14) { std::cout << i << std::endl; }
\end{lstlisting}
\end{figure}


\textbf{Implementation.} This is implemented using AST-level arid node tagging, matching expressions.

\textbf{Soundness.} This heuristic is sound because it has the full type and
expression information available.

\subsubsection{Collection Size}
The size of a collection cannot be a negative number, so when comparing the length of a container to zero, some mutants resulting from the comparison may produce unreachable code and make for unproductive test goals.

\begin{figure}[H]
\begin{lstlisting}[language=go,style=diffs]
if len(l) |>| 0:
  return l[1]
\end{lstlisting}
\vfill\hrule\vfill
\begin{lstlisting}[language=go,style=diffs]
if len(l) H<H 0:
  return l[1]
\end{lstlisting}
\end{figure}

The same also holds for collections in other languages, although it is not always easy to detect when the length is accessed. In Java, the length method can be detected for all the standard library collections by checking the inheritance chain. In Go and Python, the len builtin function can be detected with ease, and for C++, the size method can be checked for, along with
iterators or inheritance chain.

\textbf{Implementation.} This is implemented using AST-level arid node tagging, matching expressions.

\textbf{Soundness.} This heuristic is sound because it has the full type and
expression information available, barring the redefinition of a \texttt{len}
function in Python or hotplugging a patched class in Java standard library.

\subsubsection{Trivial Methods} 
Most programming languages have different types of ``boilerplate'' code that is required, but rarely considered as important to be tested by developers. For example, in Java there are methods like \texttt{equals}, \texttt{hashCode}, \texttt{toString}, \texttt{clone}, and they are usually implemented by using existing libraries like the Objects API in Java or Abseil Hash in C++. While it is possible that these methods do indeed contain bugs, the developer feedback on the productivity of corresponding mutants clearly indicates that mutants in such methods are not productive.

\begin{figure}[H]
\begin{lstlisting}[style=diffs]
@Override
public boolean equals(Object o) {
  if (|!(o instanceof CellData)|) {
    return false;
  }
  CellData that = (CellData) o;
  return Objects.equals(exp, that.exp) && Objects.equals(text, that.text);
}
\end{lstlisting}
\vfill\hrule\vfill
\begin{lstlisting}[style=diffs]
@Override
public boolean equals(Object o) {
  if (HfalseH) {
    return false;
  }
  CellData that = (CellData) o;
  return Objects.equals(exp, that.exp) && Objects.equals(text, that.text);
}
\end{lstlisting}
\end{figure}

\textbf{Implementation.} This is implemented using AST-level arid node tagging, matching expressions.

\textbf{Soundness.} This heuristic is not sound because it relies on the code
style recommendation on implementing such methods.

\subsubsection{Early Exit Optimizations}
Linus Torvalds famously states that \textit{"...if you need more than 3 levels
of indentation, you're screwed anyway, and should fix your program."} in the
kernel coding style. While this is sometimes hard to accomplish, having less
things to remember is a good thing, so it is encouraged by the code style to
return early if possible.

Consider the following mutant:

\begin{figure}[H]
\begin{lstlisting}[style=diffs]
  log.infof("network speed: %v", bytesH/Htime)
  Map<String, Integer> ExtractPrices(List<Product> products) {
-   |if (products.empty()) {|
-    |return ImmutableMap.of();|
-   |}|
    // Translation logic.
}
\end{lstlisting}
\end{figure}

The early return just makes the code easier to understand but has no effect on
the behavior, and the produced equivalent mutant is a unproductive test goal.

\textbf{Implementation.} This is implemented using AST-level arid node tagging, matching expressions.
This condition triggers when an empty container (e.g., \texttt{ImmutableMap.of()}) is returned if one of the
parameters is checked for emptiness. The checks for emptiness range from zero or
null-looking expressions, invocations of len or \texttt{size} or \texttt{empty} methods on a
container of an appropriate type that depends on the language (e.g., hash maps,
lists, dictionaries, trees, stacks, etc.). The empty container criterion checks
for standard library containers, commonly used libraries and internal
specialized container implementations.

\textbf{Soundness.} This heuristic is not sound because the mutant might not be
equivalent.

\subsubsection{Equality and Equivalence}
Some languages have equality (\texttt{==}) and equivalence (\texttt{===})
comparison operators, where one checks whether the values look the same versus
are the same. The equivalence operators check for strict equality of both type
and value, while the standard equality is not strict and applies type coercion
and then compares values, making a string \texttt{'77}' equal to an integer
\texttt{77}, because the string gets coerced to integer. The overwhelming
feedback points that strict-to-nonstrict mutants and vice versa make for unproductive
test goals.

\begin{figure}[H]
\begin{lstlisting}[language=go,style=diffs]
if (value |===| CarType.ECO)
\end{lstlisting}
\vfill\hrule\vfill
\begin{lstlisting}[language=go,style=diffs]
if (value H!=H CarType.ECO)
\end{lstlisting}
\end{figure}

To avoid dogmatic debates, \texttt{==} is only mutated to \texttt{!=} and
\texttt{===} only to \texttt{!==}.

\textbf{Implementation.} This is implemented using AST-level arid node tagging, matching binary operators.

\textbf{Soundness.} This heuristic is not sound because it relies on the code
style recommendation on comparison operators.

\subsubsection{Acceptable Bounds}
Gating a computed result into an acceptable bound by using Math.min, Math.max,
or constrainToRange of Ints, Longs, and friends is by design unlikely to change
behavior when one of the inputs is mutated. This is similar to
the Insensitive arguments heuristic, and resulting mutants are usually unproductive.

\begin{figure}[H]
\begin{lstlisting}[language=go,style=diffs]
long newCapacity = Math.min(Math.max(|(data.length * 2L)|, minCapacity), MAX_BUFFER_SIZE);
\end{lstlisting}
\vfill\hrule\vfill
\begin{lstlisting}[language=go,style=diffs]
long newCapacity = Math.min(Math.max(H(-(data.length * 2L))H, minCapacity), MAX_BUFFER_SIZE);
\end{lstlisting}
\end{figure}

\textbf{Implementation.} This is implemented using AST-level arid node tagging, matching expressions.

\textbf{Soundness.} This heuristic is not sound; it can suppress productive
mutants that can result from mathematical operations.

\subsection{JavaScript}
\subsubsection{Closure}
Closure provides a framework for library management and module registration and
exporting. These are function calls but their semantics are for the compiler at
the language level, and mutants in nodes containing them make for unproductive test
goals.

\begin{figure}[H]
\begin{lstlisting}[language=go,style=diffs]
-  |goog.requireType('goog.dom.TagName');|
\end{lstlisting}
\end{figure}

Additional issues arise from the fact that the tests are executed in a different
environment than the final compiled obfuscated minimized optimized code, where
calls to these functions are potentially removed, replaced or modified.

\textbf{Implementation.} This is implemented using AST-level arid node tagging, matching expressions.

\textbf{Soundness.} This heuristic is sound because it has the full type and
expression information available.

\subsubsection{Annotations}
A special case of the declaration heuristic is based on JavaScript's JSDoc method of signaling implicit match and
interface types, for example, \texttt{@interface} annotations. These are variables
specially tagged in comments, and require special handling compared to other
languages where interfaces are first class citizen of the language.

\begin{figure}[H]
\begin{lstlisting}[style=diffs]
-  |/**|
-  | * @interface|
-  | */|
   apps.action.Action = function() {};
\end{lstlisting}
\end{figure}

\textbf{Implementation.} This is implemented using AST-level arid node tagging, matching expressions.

\textbf{Soundness.} This heuristic is sound because it has the full type and expression information available.

\subsection{Java}
\subsubsection{System \& Runtime Classes}
Mutants around the System and Runtime class that is used for interacting with the operating system usually produce mutants that are not good test goals. This is a special case of the Low Level APIs heuristic.

\begin{figure}[H]
\begin{lstlisting}[language=go,style=diffs]
-  |System.gc();|
\end{lstlisting}
\end{figure}

\begin{figure}[H]
\begin{lstlisting}[language=go,style=diffs]
-  |Runtime.getRuntime().exec("rm -rf " + dirName);|
\end{lstlisting}
\end{figure}

\textbf{Implementation.} This is implemented using AST-level arid node tagging, matching expressions.

\textbf{Soundness.} This heuristic is not sound; it can suppress productive
mutants.

\subsubsection{Dependency Injection Modules}

Java frequently uses annotation-based
automated dependency injection such as Guice or Dagger. Modules provide bindings for injecting implementations or constants, and usually the tests will override the production modules and register testing doubles (fakes, mocks or test implementations), so changing the production module often has no effect on the tests because the tests override the setup. Such mutants are unproductive testing goals.

\begin{figure}[H]
\begin{lstlisting}[language=go,style=diffs]
-  |bindAsSingleton(binder, CarType.ECO, EcoImpl.class);|
\end{lstlisting}
\end{figure}

\textbf{Implementation.} This is implemented using AST-level arid node tagging, matching expressions.

\textbf{Soundness.} This heuristic is not sound; it assumes that all automated
dependency injection is overridden by tests.

\subsection{Python}
\subsubsection{Main}
Python's main entry point of a program is usually an if condition checking that
the script is being invoked, and not imported by another script:

\begin{figure}[H]
\begin{lstlisting}[language=go,style=diffs]
if __name__ |==| '__main__':
  app.run()
\end{lstlisting}
\vfill\hrule\vfill
\begin{lstlisting}[language=go,style=diffs]
if __name__ H!=H '__main__':
  app.run()
\end{lstlisting}
\end{figure}

Mutants in that expression are not a good test goal.

\textbf{Implementation.} This is implemented using AST-level arid node tagging, matching expressions.

\textbf{Soundness.} This heuristic is sound, barring manipulation of \texttt{\_\_name\_\_} global.

\subsubsection{Special Exceptions}
In Python, exceptions like \texttt{ValueError} imply a programming defect, something a
compiler might catch if one was employed, not something for what a test should
be written. In that case, Python's type system would be testable in each
function by calling the function with all possible types and asserting that the
interpreter works correctly; this is not a good test goal. The \texttt{AssertionError}
should usually mean that the code is unreachable. Another special case is a
virtual method that raises \texttt{NotImplementedError} and is annotated by
\texttt{abc.abstractmethod}.

\begin{figure}[H]
\begin{lstlisting}[language=go,style=diffs]
  @abstractmethod
  def virtual_method(self):
-   |raise NotImplementedError()|
\end{lstlisting}
\vfill\hrule\vfill
\begin{lstlisting}[language=go,style=diffs]
@abstractmethod
def virtual_method(self):
  HpassH
\end{lstlisting}
\end{figure}

\textbf{Implementation.} This is implemented using AST-level arid node tagging, matching expressions.

\textbf{Soundness.} This heuristic is not sound, because it relies on the
consistent usage of control flow mechanisms.

\subsubsection{Version Checks}
Python has two major versions, namely 2 and 3, and code can be written to work
for both interpreters and language specifications. The version can be determined
by reading \texttt{sys.version\_info}. Mutants in those lines make for unproductive test goals.

\begin{figure}[H]
\begin{lstlisting}[style=diffs]
if |sys.version_info[0] < 3|:
  from urllib import quote
else:
  from urllib.parse import quote
\end{lstlisting}
\vfill\hrule\vfill
\begin{lstlisting}[style=diffs]
if @False@:
  from urllib import quote
else:
  from urllib.parse import quote
\end{lstlisting}
\end{figure}

\textbf{Implementation.} This is implemented using AST-level arid node tagging, matching expressions.

\textbf{Soundness.} This heuristic is not sound, because a productive mutant
could conceivably appear in version detection code.

\subsubsection{Multiple Return Paths}
The code style requires Python programs to explicitly return None in all leafs
if there are multiple return statements: it forbids the explicit return None
that Python would return when there is no return statement in some path.
Removing those return statements does not make for a good test goal.

\begin{figure}[H]
\begin{lstlisting}[style=diffs]
  log.infof("network speed: %v", bytesH/Htime)
  def GetBuilder(x):
    if x < 10:
      logging.info('too small, ignoring')
-     |return None|
    elsif x > 100:
       return LargeBuilder()
    else:
      return SmallBuilder()
\end{lstlisting}
\end{figure}

\textbf{Implementation.} This is implemented using AST-level arid node tagging, matching complex code structures. The triggering condition is that all leaf nodes are a return
statement.

\textbf{Soundness.} This heuristic is not sound, because it relies on the code
style recommendation.

\subsubsection{Print}
In Python2, print is a first-class citizen of the AST; it is not a function that
is called using a \texttt{CallExpr}(call expression, e.g. function or method invocation). While this is covered by the Low Level API
heuristics, it is worth noting that Python requires handling this differently.

\begin{figure}[H]
\begin{lstlisting}[style=diffs]
-  |print 'exiting...'|
\end{lstlisting}
\end{figure}

\textbf{Implementation.} This is implemented using AST-level arid node tagging, matching expressions.

\textbf{Soundness.} This heuristic is not sound.

\subsection{Go}
\subsubsection{Memory Allocation}

Go has a built-in make function to allocate and initialize objects of type
\texttt{slice}, \texttt{map} or \texttt{chan}. The size parameter is used for specifying the slice
capacity, the map size or the channel buffer capacity. The initial capacity will
be grown by the runtime as needed, so changing it is undetectable by functional
tests. This is a special case of the generic memory and capacity functionality,
but it is worth explicitly mentioning because of the builtin status of this
function and the AST handling.

\begin{figure}[H]
\begin{lstlisting}[language=go,style=diffs]
buf := make([]byte, 4, |4+3*10|)
\end{lstlisting}
\vfill\hrule\vfill
\begin{lstlisting}[language=go,style=diffs]
buf := make([]byte, 4, H4+3/10H)
\end{lstlisting}
\end{figure}

\textbf{Implementation.} This is implemented using AST-level arid node tagging, matching expressions.

\textbf{Soundness.} This heuristic is sound, since it relies on full expression
and type information. Suppressed mutants are functionally equivalent.

\subsubsection{Statement Deletion} Go has a strict opinionated compiler, and unlike
most others, it has very few flags that can affect the behavior. For example, including an
unused package is a compiler error, and defining an unused identifier is also a compiler
error. In C++, it is easy to pass a flag to gcc or clang to make this only a
warning, whereas in Go that is impossible. Deleting statements or blocks of
statements almost invariably produces unbuildable code and the mutant appears
killed because the test fails (to build). There is a way to work around this,
that is employed when deleting Go statements.  First, the statement under
deletion is traversed by a recursive AST visitor, and all symbols that are used
are recorded. This includes included package literals, variables and functions,
but excludes types and built-in functions. Once the list of used symbols is
computed, the deletion can proceed, in a form of a replacement: everything that
was used in the statement under deletion is put into an unnamed slice of type
\texttt{[]interface{}}. While this is a ``hack'', this is the only way to
delete code without semantically analyzing the rest of the translation unit,
which then introduces many issues with byte offsets.

\begin{figure}[H]
\begin{lstlisting}[style=diffs]
-  |var v []string|
\end{lstlisting}
\vfill\hrule\vfill
\begin{lstlisting}[style=diffs]
H_ = []interface{}{v}H
\end{lstlisting}
\end{figure}

\begin{figure}[H]
\begin{lstlisting}[style=diffs]
v := "-42"
-  |i, err := strconv.Atoi(v)|
\end{lstlisting}
\vfill\hrule\vfill
\begin{lstlisting}[style=diffs]
v := "-42"
H_ = []interface{}{strconv.Atoi, v}H
\end{lstlisting}
\end{figure}

\textbf{Implementation.} This is implemented using AST-level arid node tagging, matching complex expressions. The deleted code is recursively visited by a custom AST visitor that
collects information about variables and functions referenced and extracts the
full list of symbols that are referenced therein. The replacement slice is
constructed from all eligible objects.

\textbf{Soundness.} This heuristic is sound in a sense that it will produce
compilable code, since it relies on full expression and type information. It
does not suppress mutants.

\fi

\end{document}